\documentclass[11pt]{article}
\usepackage{graphicx}
\usepackage{caption}
\usepackage{subcaption}
\usepackage{amsmath}
\usepackage{hyperref}
\usepackage{authblk} 
\usepackage[square,numbers]{natbib}
\usepackage{tabularx}
\usepackage{adjustbox}
\usepackage{booktabs}

\usepackage{longtable}  

\begin{document}

\title{Machine Learning Identification of Gravitationally Microlensed Gamma-Ray Bursts}

\author[1]{Mohammad H. Zhoolideh Haghighi\thanks{zhoolideh@kntu.ac.ir}}
\author[2]{Zeinab Kalantari\thanks{zeinab.kalantari70@gmail.com}}
\author[3, 4]{Sohrab Rahvar}
\author[5]{Alaa Ibrahim}

\affil[1]{Department of Physics, K.N. Toosi University of Technology, Tehran P.O. Box 15875-4416, Iran}
\affil[2]{Research Center for High Energy Physics, Sharif University of Technology, Tehran 11365-9161, Iran}
\affil[3]{Department of Physics, Sharif University of Technology, Tehran 11365-9161, Iran}
\affil[4]{Research Center for High Energy Physics, Sharif University of Technology, Tehran 11365-9161, Iran}
\affil[5]{Department of Physics, College of Science, P.O.Box 36, P.C. 123, Muscat, Sultanate of Oman}

\maketitle

\begin{abstract}
Gravitational microlensing of gamma-ray bursts (GRBs) provides a unique opportunity to probe compact dark matter and small-scale structures in the universe. However, identifying such microlensed GRBs within large datasets is a significant challenge. In this study, we develop a machine learning approach to distinguish Lensed GRBs from their Non-Lensed counterparts using simulated light curves. A comprehensive dataset was generated, comprising labeled light curves for both categories. Features were extracted using the Cesium package, capturing critical temporal properties of the light curves. Multiple machine learning models were trained on the extracted features, with Random Forest achieving the best performance, delivering an accuracy of 86\% and an F1 score of 0.86 (0.87) for the Non-Lensed (Lensed) class. This approach successfully demonstrates the potential of machine learning for identifying gravitational lensing in GRBs, paving the way for future observational applications.
\end{abstract}


\section{Introduction}

Gamma-ray bursts (GRBs) are among the most luminous transient events in the universe, originating from phenomena such as the collapse of massive stars or the merger of compact objects \citep{meszaros2006gamma, kumar2015physics}. Their fleeting nature and diverse temporal structures provide a wealth of information about the physical processes at play, but also pose significant challenges for analysis.
Given the cosmological distances of GRBs, gravitational lensing offers a unique opportunity to explore the compact small-scale structures of the Universe, potentially revealing information about compact dark matter objects  \citep{oguri2019strong, Kalantari}. Also, it provides a new tool for studying the cosmological models\citep{Kalantari3}.
Theoretical studies on gravitational lensing of GRBs have been conducted for decades  \citep{Paczynski86}. It has been proposed that GRBs can be gravitationally lensed, producing multiple images of the same burst. These images are observed as light curves arriving at different times due to variations in the light paths. In strong lensing, the time delay between images exceeds the burst duration. However, in the subclass of microlensing, the time delay between the two rays might be longer than the burst duration, causing the lensed images to appear as echoes within the same light curve. These echoes manifest as similar local peaks separated by the time delay, corresponding to unresolved lensed images  (\cite{3sim,Population3,Ji2018StrongLO,Kalantari,Kalantari2}).
In this study, we utilized long gamma-ray bursts with durations of $T_{90} \geq 2 $ seconds, observed by the Fermi Gamma-Ray Burst Monitor (GBM) (\cite{FermiInstroment}), to simulate microlensed GRBs. Our focus is on lensing effects that generate two images, which appear as local peaks (or echoes) in the light curve. These peaks exhibit identical temporal variability but differ in magnification/flux and are separated by a time delay. 

We use long GRBs (with $T_{90} \geq 2 s$) data of the Fermi Gamma-Ray Burst Monitor (GBM)  for simulating microlensed GRB (\cite{FermiInstroment}). To model this, we performed Monte Carlo simulations using a point-mass lens model for mass within the range of $10^3 - 10^7  M_{\odot}$. This mass range was chosen because our analysis targeted time delays of less than 300 seconds. The time delay for GRBs lensed by point masses is estimated to be related to the mass with the function $\Delta t= 50 s\ M_L/10^6M_{\odot}$ \citep{6sim}.

 Identifying microlensed GRBs requires efficient methods for analyzing their light curves, as the lensing-induced features are often subtle and difficult to discern.
 Machine learning (ML) has revolutionized the analysis of complex datasets in various astrophysical domains and the rapid growth of astronomical data, combined with machine learning techniques, has significantly advanced the field of astronomy (see \cite{Baron2019}, \cite{Zhoolideh2025}, \cite{Wang2024}, \cite{Ayubinia2025}, \cite{Zhoolideh2023})
  In time-domain astronomy, ML has been employed for tasks such as variable star classification \citep{richards2011}, transient detection \citep{moller2020supernnova}, and even anomaly detection in large-scale surveys \citep{lochner2016photometric}. These applications demonstrate the power of ML to extract meaningful patterns from noisy, high-dimensional data, making it a natural choice for analysing GRB light curves.

Specifically, ML has shown great promise in GRB studies. Previous works have utilised ML to classify GRB types (long versus short bursts) \citep{zhang2009discriminating}, predict redshifts \citep{krumpe2007photometric}, and estimate spectral parameters \citep{wang2020automatic}. The ability of ML to identify rare events, such as gravitationally lensed GRBs, is particularly valuable given the rarity of such phenomena and the vast amount of observational data produced by modern telescopes.

A crucial step in ML-based analysis of time-series data, such as GRB light curves, is feature extraction. The increasing volume of time-domain photometric data from large-scale surveys necessitates efficient and scalable feature extraction methods for reliable classification. The \texttt{light-curve} library, for instance, provides a comprehensive suite of features characterizing light curve morphology, magnitude distribution, and periodicity. It has already been integrated into real-time pipelines for surveys like ZTF and LSST \citep{lLavrukhina}. Similarly, \cite{Khakpash2021} extracts features from simulated microlensing light curves for classification, while \cite{Edes2021} utilizes boosted neural networks, decision trees, and various feature selection techniques, including wavelet decomposition and neural network-based ranking, for the multiclass classification of PLAsTiCC light curves spanning 14 object types. Moreover, the analysis of X-ray light curves from Swift observations has revealed internal plateau features in a subset of GRBs, providing valuable insights into the properties of nascent magnetars \citep{Lyons2010}.

Automated tools like the \texttt{Cesium} package facilitate the extraction of temporal and frequency domain features, which are vital for distinguishing between Lensed and non-Lensed GRBs \citep{naul2018cesium}. These features, when fed into robust classification algorithms such as Random Forests, can substantially improve the identification of gravitationally microlensed events.

In this study, we investigate whether the presence of two distinct peaks in a GRB light curve can be attributed to gravitational lensing. 
To identify promising microlensed GRB candidates, a thorough analysis of the light curve's physical properties across different energy bands -particularly the time separation between two peaks and their relative brightening in various energy ranges \citep{Kalantari}- is essential. However, as a first step—and given the limitations of our current dataset—we employ machine learning to detect potential candidates based on morphological features alone.
To this end, we generate two mock datasets—each consisting of 2000 light curves—representing Lensed and non-Lensed GRBs. Features capturing both statistical and physical characteristics are extracted using \texttt{Cesium}, and multiple ML models are trained and evaluated on this feature set. Among the tested classifiers, the Random Forest algorithm achieves the best performance, with an accuracy of 86\% and an F1 score of 0.87, demonstrating its effectiveness in detecting gravitationally microlensed GRBs.

This paper is structured as follows: In Section~\ref{gr_lesned}, we explain the physics of gravitational microlensing of GRBs. In Section~\ref{data}, we describe the simulation of GRB light curves, the creation of our training and test sets, and feature extraction. Section~\ref{model} explains the machine learning methodologies employed. Section~\ref{sec:results} presents the classification results, compares the performance of different models, and discusses the implications of our findings. Finally, in Section \ref{sec:application}, we apply our trained models on real observed GRBs.


\section{Gravitatioal Microlensing}
\label{gr_lesned}
Gravitational lensing occurs when a massive object is positioned near the line of sight between an observer and a source. A point mass gravitational lens amplifies the source and produces two distinct images. In this study, a GRB serves as the source, and each image, influenced by gravitationally induced time delay and varying magnification, can be identified through the observed burst light curve.
\par
Two key parameters in gravitational lensing play a crucial role in analyzing the GRB light curve. The first is the time delay between the two images, determined under the assumption of a point mass model for the lens. By defining the normalized impact parameter as 
$y=\frac{\beta}{\theta_E}$ where $\beta$ represents the angular source position and $\theta_E$ the Einstein radius. The time delay can be calculated as described by \citep{Kalantari}.

\begin{equation}
\Delta t= t^+ - t^-= 4\frac{GM_L}{c^3} (1+z_L)[\frac{y}{2}\sqrt{y^2+4}+\ln(\frac{\sqrt{y^2+4}+y}{\sqrt{y^2+4}-y})],
\label{eq:delay}
\end{equation}
Here, $M_L$ represents the lens mass, while $\Delta t = t^- - t^+$ denotes
The relative time delay between fainter (negative parity) and brighter (positive parity) lensed images.
The lens redshift is given by $z_L$.
\par
The second key parameter is the flux ratio between the two images, which is determined by the ratio of their respective magnification factors \citep{Kalantari}.
\begin{equation}
R=|\frac{\mu_+}{\mu_ -}|=\frac{y^2+2+y \sqrt{y^2+4}}{y^2+2-y \sqrt{y^2+4}},
\label{eq:magnification}
\end{equation}
Here,$\mu_{\pm}$ represents the magnification factors of the positive and negative parity images.
The time delay $\Delta t$ is positive when $t^- \geq t^+$, meaning the
 fainter peak of the light curve, which corresponds to the image with the lower magnification, will arrive later than the brighter one, which corresponds to the image with the higher magnification ({\it i.e.} $t^+ \leq t^-$) as derived from Eqs. (\ref{eq:delay}) and (\ref{eq:magnification}).
 \par
Assuming a lens mass of $5 \times 10^5 M_\odot$ and placing the lens at a cosmological redshift of approximately half the GRB's distance ($z_L=1\sim1Gpc$), the angular separation between the two images, $\|\theta_+-\theta_-\|$, would be on the order of $10^{-2}$ arcseconds.
The angular resolution of Fermi/GBM ($\sim 2^\circ$) is insufficient to distinguish between two gravitationally lensed images with small angular separations. However, lensing events with closely spaced images and shorter time delays can create detectable repeated structures or echoes with different magnifications within the same GRB light curve.
As a result, the two gravitationally lensed images of a GRB would be observed as a single event, with the images manifesting as consecutive peaks or sub-bursts within the same light curve. This occurs when the time delay between the two images falls within the range where it is shorter than the total observation period but longer than the burst duration.

\section{Data} \label{data} 
We used data from the GBM instrument of the Fermi telescope to obtain the light curves of GRBs. Fermi/GBM records GRBs in a broad energy range (8 keV to 40 MeV) with a field of view of $\geq 8$ sr and has twelve thallium-activated sodium iodide (NaI) and two bismuth germanate (BGO) detectors (\cite{FermiInstroment}). For simulation, we extracted light curve data  from the NaI detectors with the highest signal using the time-tagged event (TTE) files and considered long GRBs (with $T_{90} \geq 2 s$) 
 and then we subtract the background for a zeroth-order polynomial model (a constant) is chosen to fit the background. For
 performing a Monte Carlo simulation of mock lensed GRBs, we select 81 long GRBs from the Fermi/GBM catalog \citep{von_Kienlin_2020} that have one pulse in their light curves with a time bin of 64 ms. 
 In Section \ref{simulation}, we illustrate how we produced mock microlensed light curves of these GRBs by imposing a time delay and a magnification factor for the lensed mass, resulting from solving the gravitational lensing equations mentioned in Section \ref{gr_lesned}.
In the following section, we aim to identify whether the presence of two local peaks in the light curve of a GRB is a result of gravitational lensing. To achieve this, we generate two datasets—one with mock lensed data and another with non-lensed data.

\subsection{Simulations}
\label{simulation}

While the sub-arcsecond angular separation of two images of a GRB lensed by the previously mentioned mass will not be resolved by Fermi, and the GRB will be recorded as a single event, the light curve recorded by Fermi/GBM will be a superposition of the two light curves corresponding to the two lensed images. This can be resolved when the time delay is shorter than the GRB recording time. 
\par
We aim to generate a semi-synthetic dataset that incorporates original light curves derived from real data. Using our Monte Carlo simulation, we will classify the light curves into two categories: Lensed and non-Lensed.
All light curves were constructed using photons in the full energy range of 8–150 keV, which corresponds to the most efficient detection interval of Fermi/GBM \citep{FermiInstroment}. A key characteristic of microlensed GRB light curves is the similarity between the two peaks, along with distinct differences in their temporal structure compared to non-lensed events. Previous studies \citep{Kalantari2} have investigated this using chi-squared tests, whereas in this work we employ a machine learning approach to identify distinctive features.

For the identification of robust microlensed GRB candidates in real observations, additional physical constraints, such as the magnification ratio and the time delay between peaks across different energy bands \citep{Kalantari}, must also be taken into account. However, since our aim here is to identify possible candidates, our simulations are limited to the full energy band. We therefore focus exclusively on evaluating peak similarity and the overall temporal structure of the light curves.
\par
\begin{figure*}[!htbp]
\begin{minipage}{0.47\textwidth}
\includegraphics[width=\linewidth]{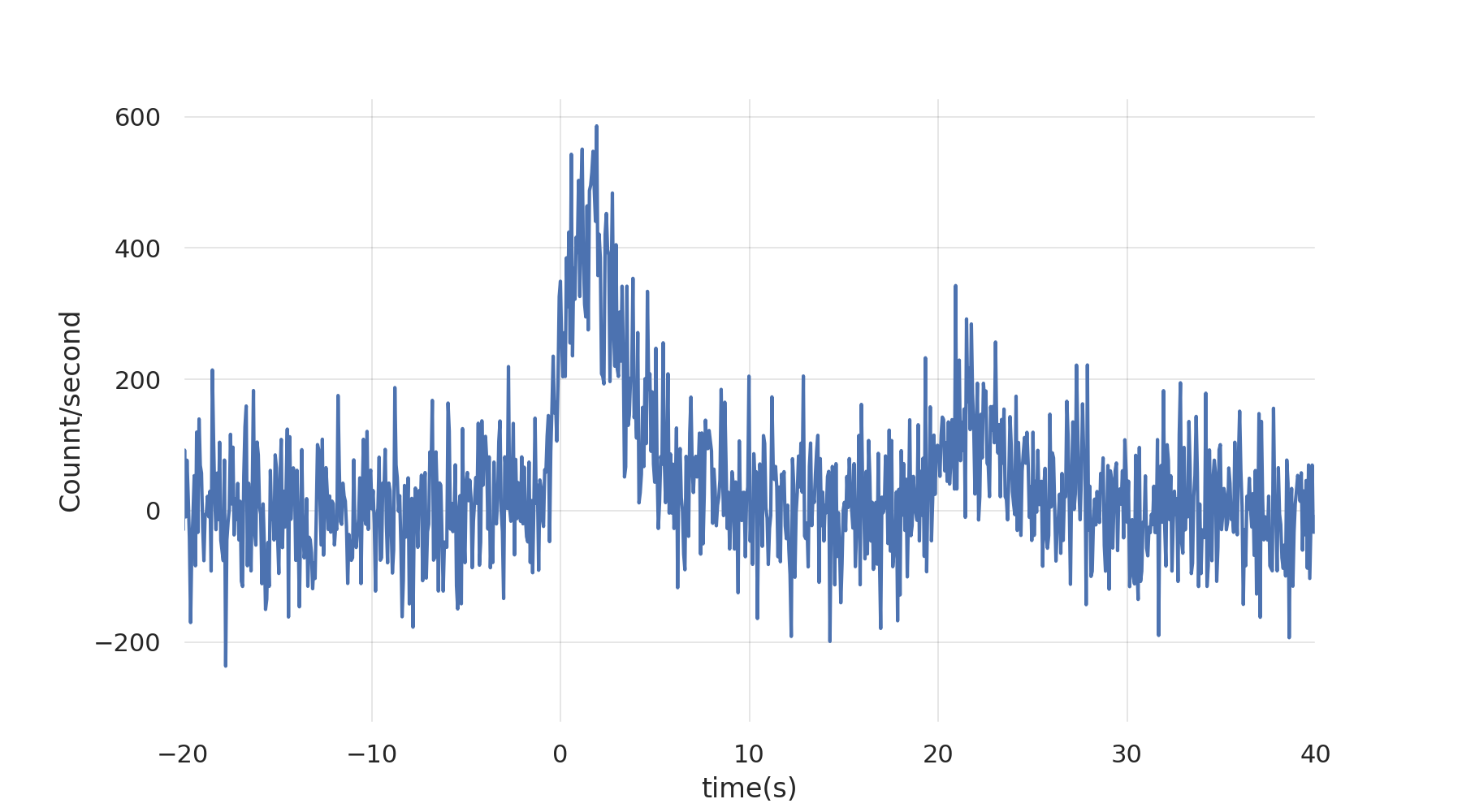}
\end{minipage}
\hfill
\begin{minipage}{0.47\textwidth}
\includegraphics[width=\linewidth]{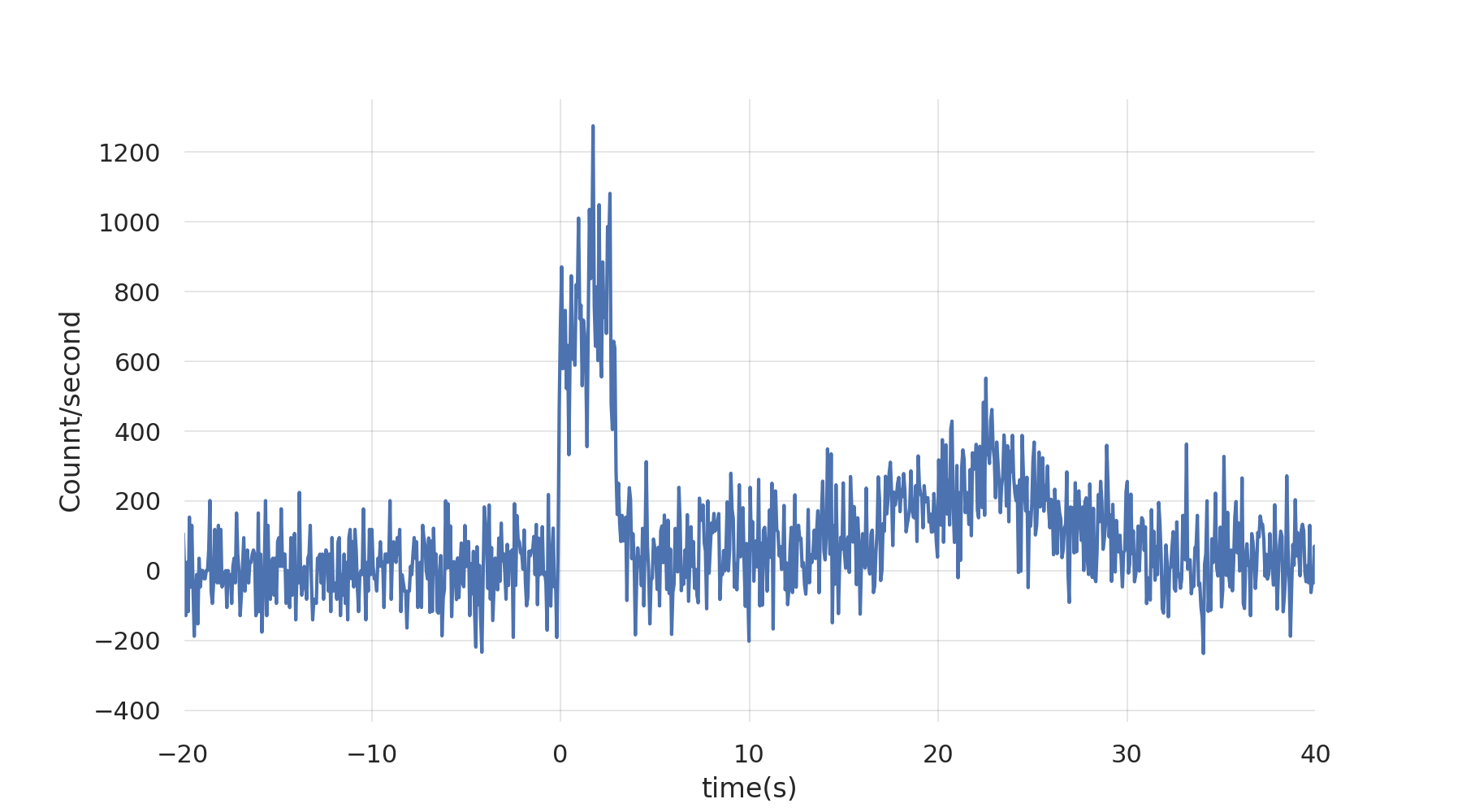}
\end{minipage}

\caption{Left: simulated lensed light curve (see Eq.~\protect\ref{eq:superimpose_lensed}); Right: simulated non-lensed light curve (see Eq.~\protect\ref{eq:superimpose_not_lensed}).}

\label{fig:simulated_lensed}
\end{figure*}

\begin{itemize}
    \item \textbf{Mock Lensed light curve:}
Based on the physics of gravitational microlensing, for a redshifted lens mass $(M_L(1+z))$ assumed uniformly distributed over the interval $10^3 - 10^7  M_{\odot}$and a normalized impact parameter uniformly distributed within $\beta =[0,1]$, we first solve the lensing equations (Eq(\ref{eq:delay}) and Eq(\ref{eq:magnification})) to find time delay $\Delta t$ and a magnification ratio $R$.
Then we simulated a Lensed light curve by superimposing the count rate $I(t)$ of a real GRB with itself as shown in Fig.\ref{fig:simulated_lensed}(Left), by imposing a time delay $\Delta t$ and a magnification ratio $R$. The resulting Lensed light curve count rate is
\begin{equation}
I_{Lensed}(t)=\frac{R}{R+1} I(t) + \frac{1}{R+1}I(t+\Delta t).
\label{eq:superimpose_lensed}
\end{equation}

 The crucial feature of simulating Lensed light curves is that two local pulses in the GRB light should have the same temporal profile, separated by a specific time interval. So both count rates of the simulated Lensed light curve are chosen from the same GRB light curve.
 
    \item \textbf{Mock Non-Lensed light curve:}
    For the Non-Lensed simulation, we generate a light curve pattern with two local peaks, where the fainter peak occurs later than the brighter one, but each peak exhibits a distinct temporal profile. Therefore, we superimpose the count rate $I(t)$ of a real GRB (for the brighter peak) with another GRB's light curve 
$I'(t)$ (for the second peak), as shown in Fig.\ref{fig:simulated_lensed} (right). This is done by introducing a randomly chosen time delay 
$\Delta t$ and an amplification factor,$A$, to ensure that the fainter peak appears later than the brighter one. The resulting light curve count rate is:
\begin{equation}
I_{Non-lensed}(t)= A I(t) + I'(t+\Delta t).
\label{eq:superimpose_not_lensed}
\end{equation}
The main feature of the simulation of the Non-Lensed light curve is that the two pulse shapes are different, and we choose the second peak data from another GRB light curve. 

Based on the method explained above, we simulate 4000 light curves, including 2000 Lensed and 2000 Non-Lensed GRBs. We extract the important features of the generated light curves below, which will serve for training the machine learning models. 

\end{itemize}

\subsection{Feature Selection and Extraction}
\label{Feature_Selection}

To detect gravitational microlensed gamma-ray bursts (GRBs), we extracted a set of statistical and temporal features from the light curves of GRBs. These features were chosen based on their ability to capture the underlying characteristics of the light curves, which are critical for distinguishing microlensed GRBs from non-microlensed ones. The features were computed using the \texttt{cesium} feature extraction package \citep{cesium}, which provides a comprehensive suite of tools for time-series analysis in astrophysics.
The following features were extracted from the light curves:
\begin{itemize}
    \item \textbf{Amplitude}: Measures the range of variability in the light curve count rate.
    \item \textbf{Percent Beyond 1 Std}: Quantifies the fraction of data points beyond one standard deviation from the mean.
    \item \textbf{Maximum, Minimum, and Median}: Capture the extreme values and central tendency of the light curve.
    \item \textbf{Skew}: Measures the asymmetry of the light curve distribution.
    \item \textbf{Maximum Slope}: This feature quantifies the steepest slope observed in the light curve, which is indicative of rapid changes in brightness. 
    
    \item \textbf{Percentage Beyond 1 Standard Deviation}: This feature measures the proportion of data points in the light curve that lie beyond one standard deviation from the mean. 

    \item \textbf{Standard Deviation (Std)}: Represents the variability in the light curve.
    \item \textbf{Weighted Average}: Computes the mean weighted by measurement uncertainties.
    \item \textbf{Anderson-Darling and Shapiro-Wilk}: Statistical tests for normality, which help identify deviations from Gaussian behavior.
    \item \textbf{Stetson J and K}: Measures of variability and correlation in time-series data \citep{stetson1996}.
    \item \textbf{Median Absolute Deviation (MAD)}: A robust measure of dispersion.
    \item \textbf{Percent Close to Median}: Quantifies the fraction of data points near the median value.
    \item \textbf{Period Fast}: Estimates the dominant periodicity in the light curve.
    \item \textbf{QSO Log Chi-Squared Features}: Derived from quasar variability studies, these features are useful for characterizing non-periodic variability \citep{macLeod2010}.
    \item \textbf{Number of Peaks and Epochs}: Captures the structure and sampling of the light curve.
    \item \textbf{All\_times\_nhist\_numpeaks}:
      Quantifies the number of local peaks in the histogram of all time values within the light curve. It provides insight into the temporal clustering of data points.
\end{itemize}

\begin{figure}[htbp]
  \centering    \includegraphics[width=1.2\textwidth]{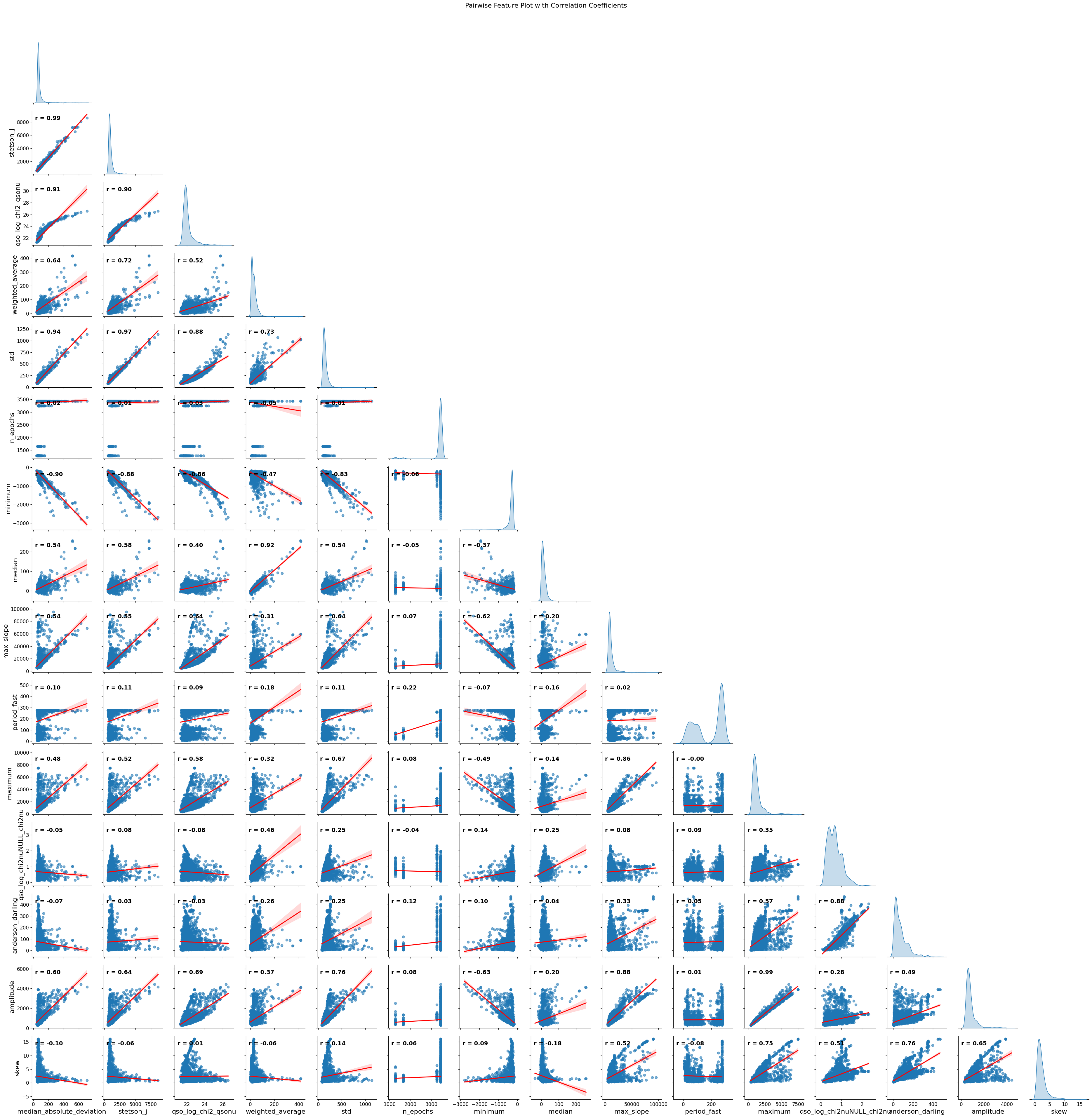}%
    \caption{Pairwise feature comparison of the extracted features. 
    The lower triangle displays scatter plots with linear regression fits (red lines) 
    and annotated Pearson correlation coefficients ($r$). 
    Diagonal panels show kernel density estimates (KDEs) representing 
    the distribution of each feature.}
    \label{fig:pair_plot}
\end{figure}

These features were selected based on their demonstrated effectiveness in characterizing astronomical time-series data \citep{richards2011, feigelson2012}. The feature extraction process was performed using the \texttt{featurize\_time\_series} function from the \texttt{cesium} package, which outputs a feature matrix $\mathbf{X}$ where each row corresponds to a GRB and each column represents a feature. To better see how the extracted features correlate, we have provided a pairwise matrix plot in Figure \ref{fig:pair_plot}. We have also included Table \ref{tab:feature_stat}, detailing all extracted features and some of their statistical features, such as mean, std and etc.

Although the features used in our machine learning model are statistical and not explicitly derived from gravitational lensing theory, they are sensitive to observable consequences of lensing, such as the distortion, duplication, and time delay between light curve components. Below, we outline the relevance of the extracted features:

 \begin{itemize}   
\item  \textbf{Statistical and Shape-Based Features:}
    Features like \texttt{amplitude}, \texttt{maximum}, \texttt{minimum}, \texttt{median}, \texttt{skew}, \texttt{std}, and \texttt{weighted\_average} describe the overall brightness and distribution of the light curve intensity. Lensing can magnify or distort the light curve profile, leading to changes in these values.

\item \textbf{Variability and Outlier Metrics:}
    Features such as \texttt{percent\_beyond\_1\_std}, \texttt{median\_absolute\_deviation}, \texttt{max\_slope}, \texttt{percent\_close\_to\_median}, and \texttt{stetson\_j}/\texttt{k} quantify the variability or irregularity of the light curves. Lensed GRBs, particularly those affected by microlensing or multiple image arrivals, may display enhanced or anomalous variability.

\item \textbf{Normality and Distribution Tests:}
    Features like \texttt{anderson\_darling} and \texttt{shapiro\_wilk} test how closely the light curve matches a Gaussian distribution. While GRB light curves are generally non-Gaussian, lensing effects can further enhance these deviations, especially by introducing multiple components.

 \textbf{Periodicity and Frequency Domain:}
    \texttt{period\_fast} can detect potential periodic or quasi-periodic behavior. In rare cases of strong lensing, repeated signals due to time delays could mimic quasi-periodic or multi-peaked structures in the light curve.

\item \textbf{QSO-based Variability Statistics:}
    \texttt{qso\_log\_chi2\_qsonu} and \texttt{qso\_log\_chi2nuNULL\_chi2nu} are adapted from quasar variability studies and reflect non-Gaussian variability patterns. Although originally designed for quasar variability studies, these statistics may capture non-Gaussian or stochastic variability patterns that could be indirectly influenced by lensing in GRBs.

\item \textbf{Light Curve Morphology:}
    \texttt{all\_times\_nhist\_numpeaks} and \texttt{n\_epochs} quantify the number of significant structures and observations. Lensed GRBs may show multiple peaks or extended profiles due to delayed image arrivals, and these features can reflect such structural complexity.

\end{itemize}
Taken together, these features serve as observable proxies that encode the photometric and temporal alterations expected from gravitational lensing, even though the model does not directly use lensing parameters (e.g., mass distribution or redshift). Although these features do not directly encode lensing physics, they serve as observable proxies that reflect the morphological and temporal signatures expected from lensing. The machine learning classifier learns from these patterns to effectively distinguish between Lensed and Non-Lensed GRBs.

The extracted features will be used as input to a machine learning model for classification. The target variable $\mathbf{y}$ consists of binary labels indicating whether a GRB is microlensed or not. We split the dataset into training and testing sets using an 80-20 split, ensuring that the model's performance could be evaluated on unseen data. It is worth mentioning that to ensure consistent scaling across the dataset, we standardised the features by removing the mean and scaling to unit variance using the StandardScaler. This transformation was applied to both the training and test sets, with the scaler fitted only on the training data to prevent data leakage.

The use of machine learning for detecting microlensed GRBs is motivated by the complexity and high dimensionality of the light curve data. Traditional methods often rely on manual inspection or simple thresholding techniques, which may fail to capture subtle signatures of microlensing \citep{oguri2019}. By leveraging a diverse set of features and advanced machine learning algorithms, our approach aims to improve the accuracy and robustness of microlensed GRB detection.

\small
\begin{longtable}{lccccccc}
\caption{Descriptive Statistics of Features. The table includes the mean, standard deviation (std), minimum (min), maximum (max), and key percentiles: 25\% (first quartile), 50\% (median), and 75\% (third quartile).}
\label{tab:feature_stat}\\
\toprule
Feature & mean & std & min & max & 25\% & 50\% & 75\% \\
\midrule
\endfirsthead

\multicolumn{8}{c}{{\bfseries Continued from previous page}} \\
\toprule
Feature & mean & std & min & max & 25\% & 50\% & 75\% \\
\midrule
\endhead

\midrule
\multicolumn{8}{r}{{Continued on next page}} \\
\endfoot

\bottomrule
\endlastfoot

amplitude & 834.01 & 526.60 & 287.86 & 4434.56 & 549.93 & 687.41 & 891.14 \\
percent\_beyond\_1\_std & 0.21 & 0.06 & 0.02 & 0.32 & 0.18 & 0.22 & 0.26 \\
maximum & 1310.48 & 938.04 & 336.58 & 7460.40 & 794.29 & 1048.09 & 1430.73 \\
median & 13.33 & 15.08 & -36.72 & 257.78 & 4.19 & 10.19 & 18.99 \\
minimum & -357.54 & 205.39 & -2789.24 & -177.39 & -356.64 & -301.14 & -269.09 \\
skew & 2.31 & 1.95 & 0.27 & 16.16 & 1.19 & 1.84 & 2.72 \\
std & 150.45 & 80.01 & 74.60 & 1139.22 & 111.29 & 131.46 & 160.19 \\
weighted\_average & 30.25 & 25.31 & -0.17 & 416.91 & 13.88 & 25.32 & 38.89 \\
max\_slope & 11418.74 & 9577.36 & 4655.37 & 95384.32 & 7408.24 & 8689.79 & 11323.14 \\
anderson\_darling & 74.36 & 68.19 & 3.37 & 471.45 & 25.11 & 54.47 & 99.86 \\
percent\_beyond\_1\_std & 0.21 & 0.06 & 0.02 & 0.32 & 0.18 & 0.22 & 0.26 \\
percent\_close\_to\_median & 0.80 & 0.11 & 0.50 & 0.99 & 0.71 & 0.80 & 0.89 \\
period\_fast & 181.74 & 102.68 & 5.85 & 274.96 & 70.68 & 253.42 & 274.96 \\
qso\_log\_chi2\_qsonu & 22.08 & 0.61 & 21.26 & 26.60 & 21.75 & 21.91 & 22.17 \\
qso\_log\_chi2nuNULL\_chi2nu & 0.67 & 0.34 & 0.08 & 2.33 & 0.41 & 0.64 & 0.88 \\
shapiro\_wilk & 0.86 & 0.11 & 0.30 & 1.00 & 0.81 & 0.89 & 0.94 \\
stetson\_j & 1017.88 & 557.15 & 580.92 & 8641.36 & 784.21 & 881.32 & 1050.37 \\
stetson\_k & 0.85 & 0.09 & 0.44 & 1.00 & 0.81 & 0.87 & 0.92 \\
all\_times\_nhist\_numpeaks & 1.02 & 0.14 & 1.00 & 2.00 & 1.00 & 1.00 & 1.00 \\
n\_epochs & 3359.47 & 382.82 & 1282.00 & 3438.00 & 3438.00 & 3438.00 & 3438.00 \\

\end{longtable}

\section{Model definition \& training} \label{model}

\subsection{Machine Learning Models}
\label{subsec:models}

To detect gravitational microlensed gamma-ray bursts (GRBs), we employed a diverse set of machine learning classifiers, each chosen for its unique strengths in handling high-dimensional data and capturing complex patterns. The models used in this study include:

\begin{itemize}
    
    \item \textbf{Random Forest (RandomForestClassifier)}: An ensemble method that constructs multiple decision trees during training and outputs the mode of the classes for classification tasks \citep{breiman2001}. Random Forest is robust to overfitting and provides feature importance rankings, which are useful for interpreting the results.

    \item \textbf{Bagging Classifier (BaggingClassifier)}: An ensemble method that combines the predictions of multiple base classifiers trained on random subsets of the data \citep{breiman1996}. Bagging reduces variance and improves generalization performance.
    
    \item \textbf{AdaBoostClassifier}:  An ensemble learning algorithm that combines multiple weak classifiers to create a strong classifier by iteratively adjusting the weights of misclassified samples \citep{freund1997decision}. This method is particularly effective in reducing bias and variance, making it suitable for various classification tasks, including astronomy and astrophysics applications.

    \item \textbf{k-Nearest Neighbors (KNeighborsClassifier)}: A non-parametric method that classifies data points based on the majority class among their \(k\) nearest neighbors \citep{cover1967}. This model is effective for capturing local patterns in the data.
    
    \item \textbf{Support Vector Classifier (SVC)}: A kernel-based method that finds the optimal hyperplane to separate classes in a high-dimensional feature space \citep{cortes1995}. SVC is particularly effective for datasets with clear margins of separation.
    
    \item \textbf{Logistic Regression}: A widely used statistical model for binary classification that predicts the probability of an instance belonging to a class using a logistic function. It is trained using maximum likelihood estimation and is known for its interpretability and efficiency \citep{hastie2009elements}. Logistic Regression often serves as a strong baseline for evaluating more complex models due to its simplicity and robustness.

\end{itemize}
The diversity of the selected models ensures that both linear and non-linear relationships in the data are captured. By comparing the performance of these models, we aim to identify the most effective approach for detecting microlensed GRBs.

\subsubsection{Model Evaluation}

In classification tasks, predictions can result in four outcomes: \textit{True Positives (TP)}, where an observation is correctly predicted to belong to a class; \textit{True Negatives (TN)}, where an observation is correctly predicted not to belong to a class; \textit{False Positives (FP)}, where an observation is incorrectly predicted to belong to a class; and \textit{False Negatives (FN)}, where an observation is incorrectly predicted not to belong to a class. These outcomes are typically organized in a confusion matrix, which categorizes predictions made on test data.

Model performance is evaluated using metrics such as \textit{accuracy}, \textit{precision}, \textit{recall}, and the \textit{F\textsubscript{1} score}. Accuracy measures the proportion of correct predictions relative to the total number of predictions:

\[
\text{Accuracy} = \frac{\text{correct predictions}}{\text{all predictions}}
\]

Precision quantifies the fraction of true positives among all predicted positives:

\[
\text{Precision} = \frac{\text{true positives}}{\text{true positives + false positives}}
\]

Recall, or sensitivity, measures the fraction of true positives identified out of all actual positives:

\[
\text{Recall} = \frac{\text{true positives}}{\text{true positives + false negatives}}
\]

The \textit{F\textsubscript{1} score} combines precision and recall into a single metric by calculating their harmonic mean, providing a balanced measure of classifier performance. A score of 1 indicates perfect classification, while lower values reflect poorer performance:

\[
F_1 = \frac{2 \cdot \text{Precision} \cdot \text{Recall}}{\text{Precision + Recall}}
\]

\hspace{2cm}

Additionally, the \textit{Receiver Operating Characteristic (ROC) curve} is a graphical tool used to assess the performance of binary classifiers. It plots the \textit{True Positive Rate (TPR)} against the \textit{False Positive Rate (FPR)} across varying classification thresholds. The \textit{Area Under the Curve (AUC)} summarizes the classifier's ability to distinguish between classes, with an AUC of 1 representing perfect discrimination and 0.5 indicating no better than random chance. ROC analysis is particularly valuable for imbalanced datasets, as it evaluates performance across different thresholds, offering a robust assessment of model efficacy \citep{Fawcett2006ROC, Powers2011ROC, Bradley1997AUC, Saito2015AUC}.

\subsection{Model Training}

\begin{figure*}[ht!]
        \begin{subfigure}{0.32\textwidth}
                \includegraphics[width=\linewidth]{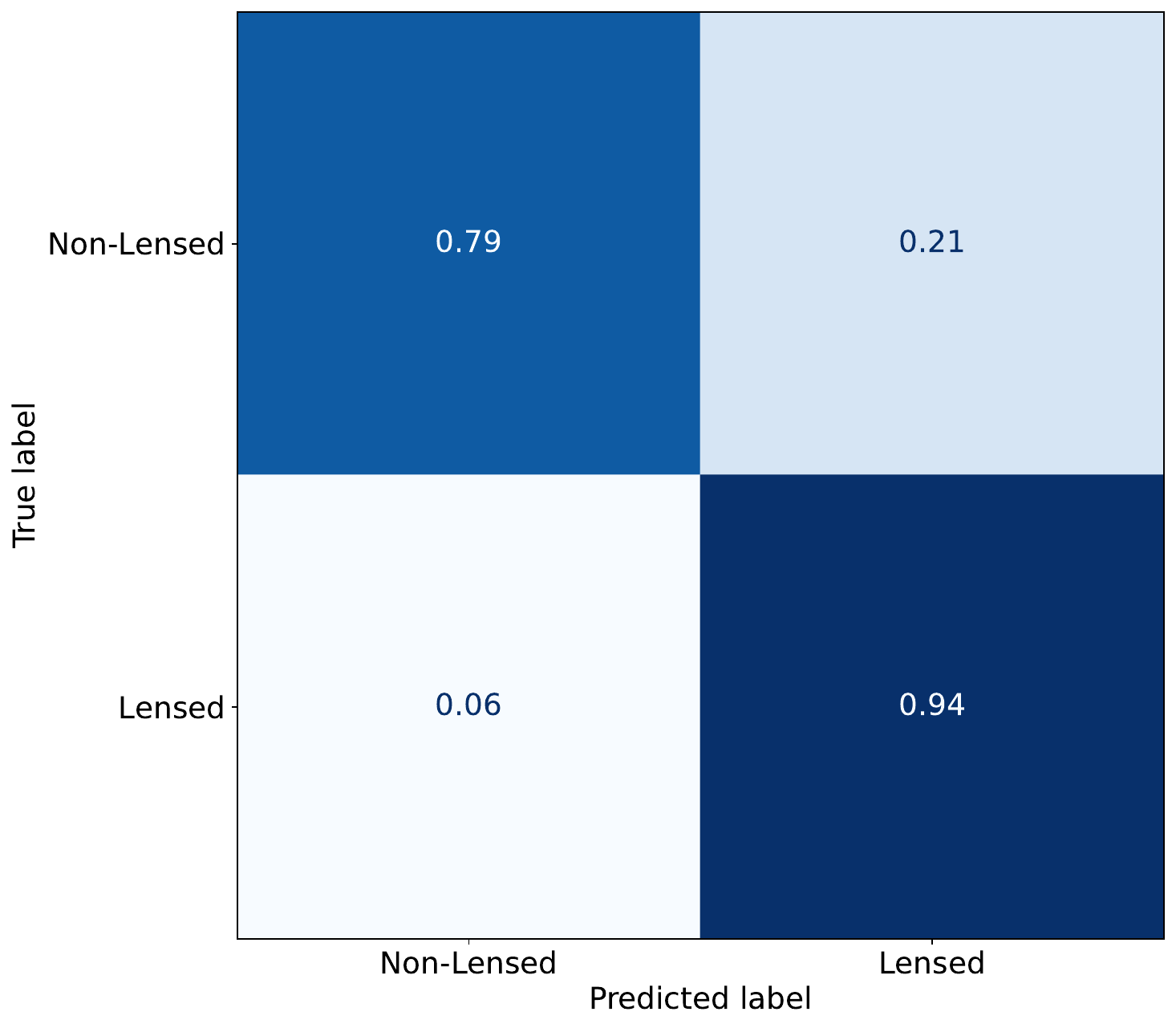}
        \caption{Random Forest}
        \label{fig:cm_rf}
    \end{subfigure}
    \hfill
    \begin{subfigure}{0.32\textwidth}
                \includegraphics[width=\linewidth]{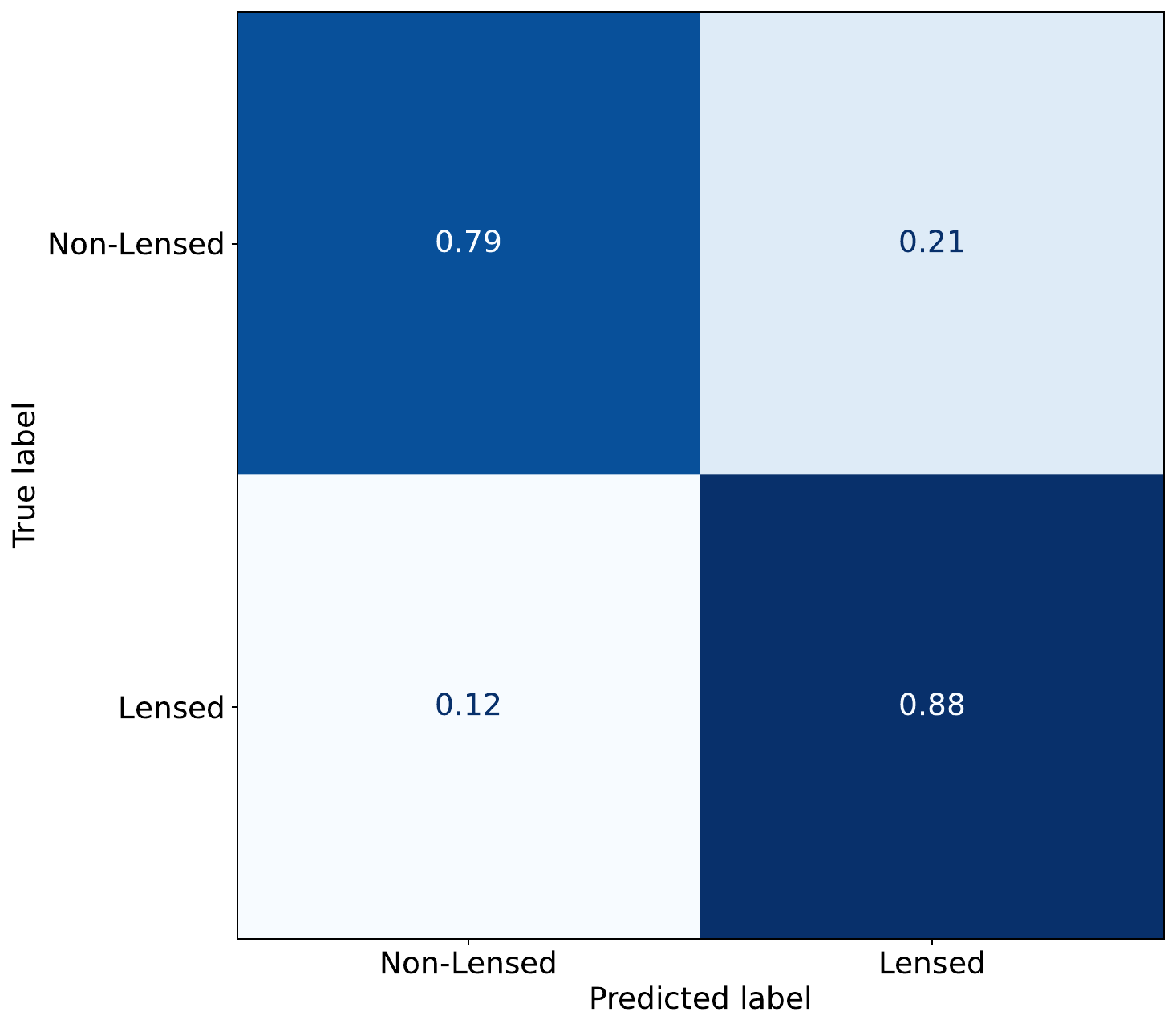}
        \caption{Bagging Classifier}
        \label{fig:cm_bagging}
    \end{subfigure}
    \hfill
    \begin{subfigure}{0.32\textwidth}
                \includegraphics[width=\linewidth]{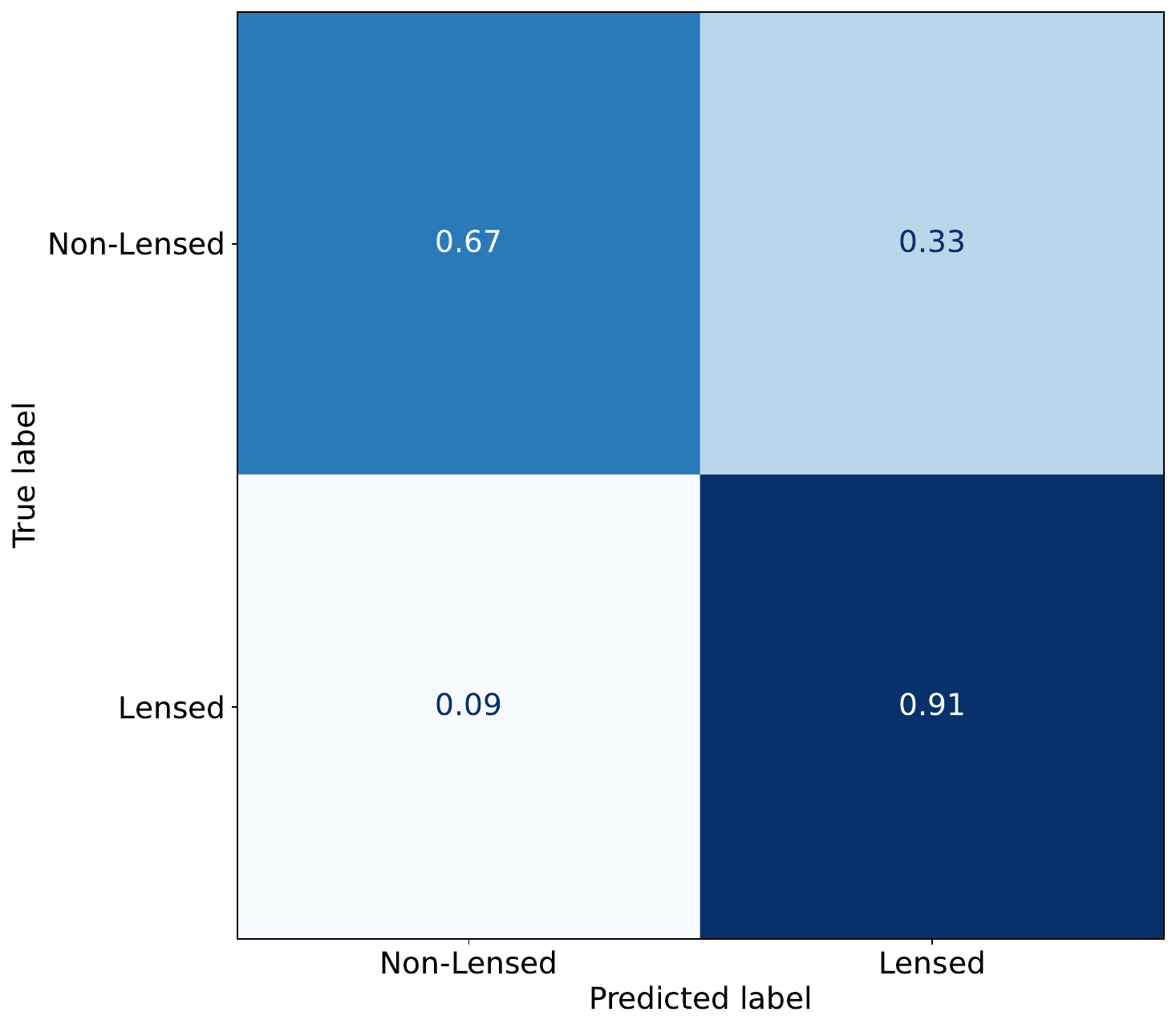}
        \caption{k-Nearest Neighbors}
        \label{fig:cm_knn}
    \end{subfigure}

    \vspace{0.5cm}

    \begin{subfigure}{0.32\textwidth}
                \includegraphics[width=\linewidth]{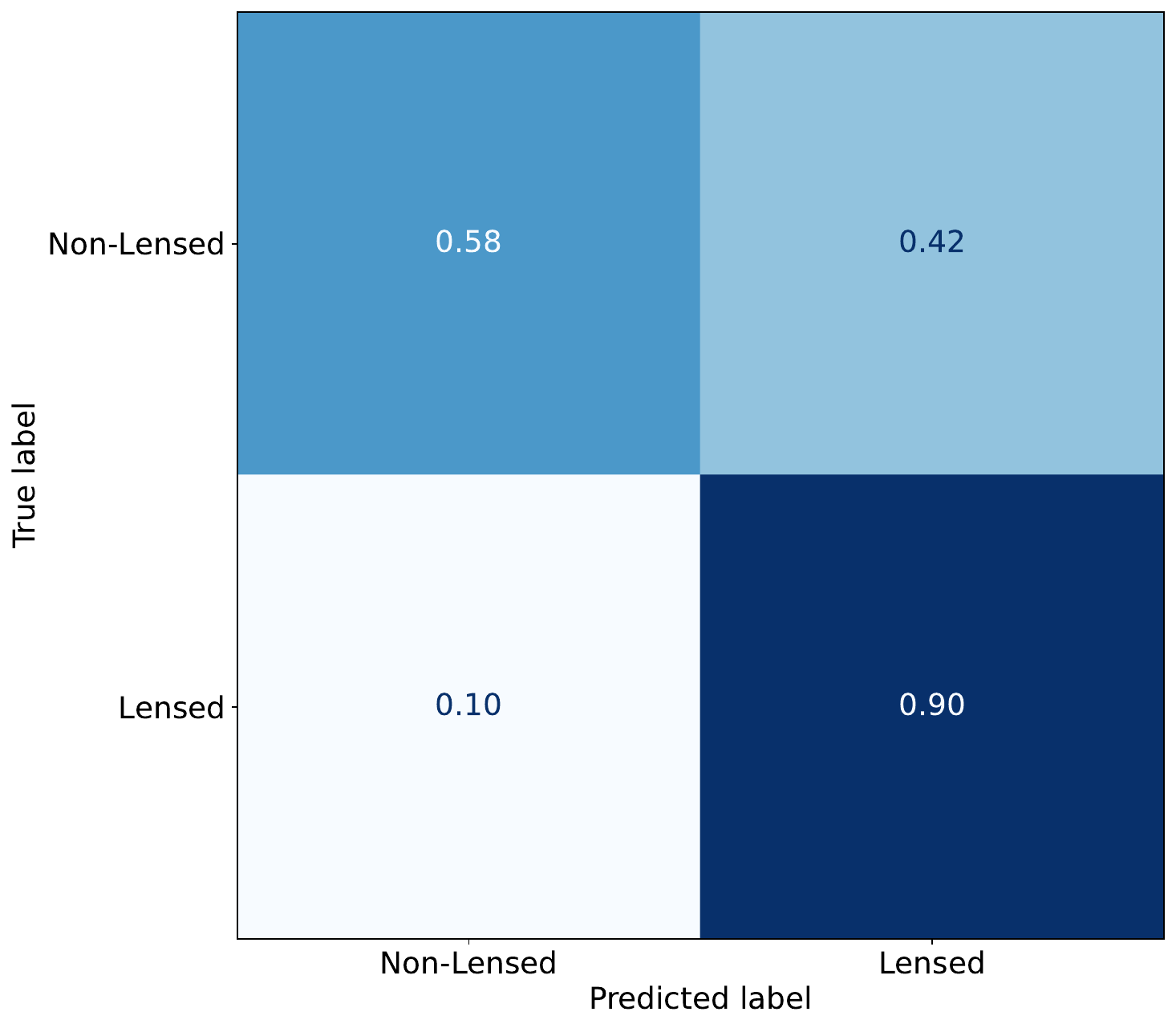}
        \caption{Support Vector Machine}
        \label{fig:cm_svm}
    \end{subfigure}
    \hfill
    \begin{subfigure}{0.32\textwidth}
        \includegraphics[width=\linewidth]{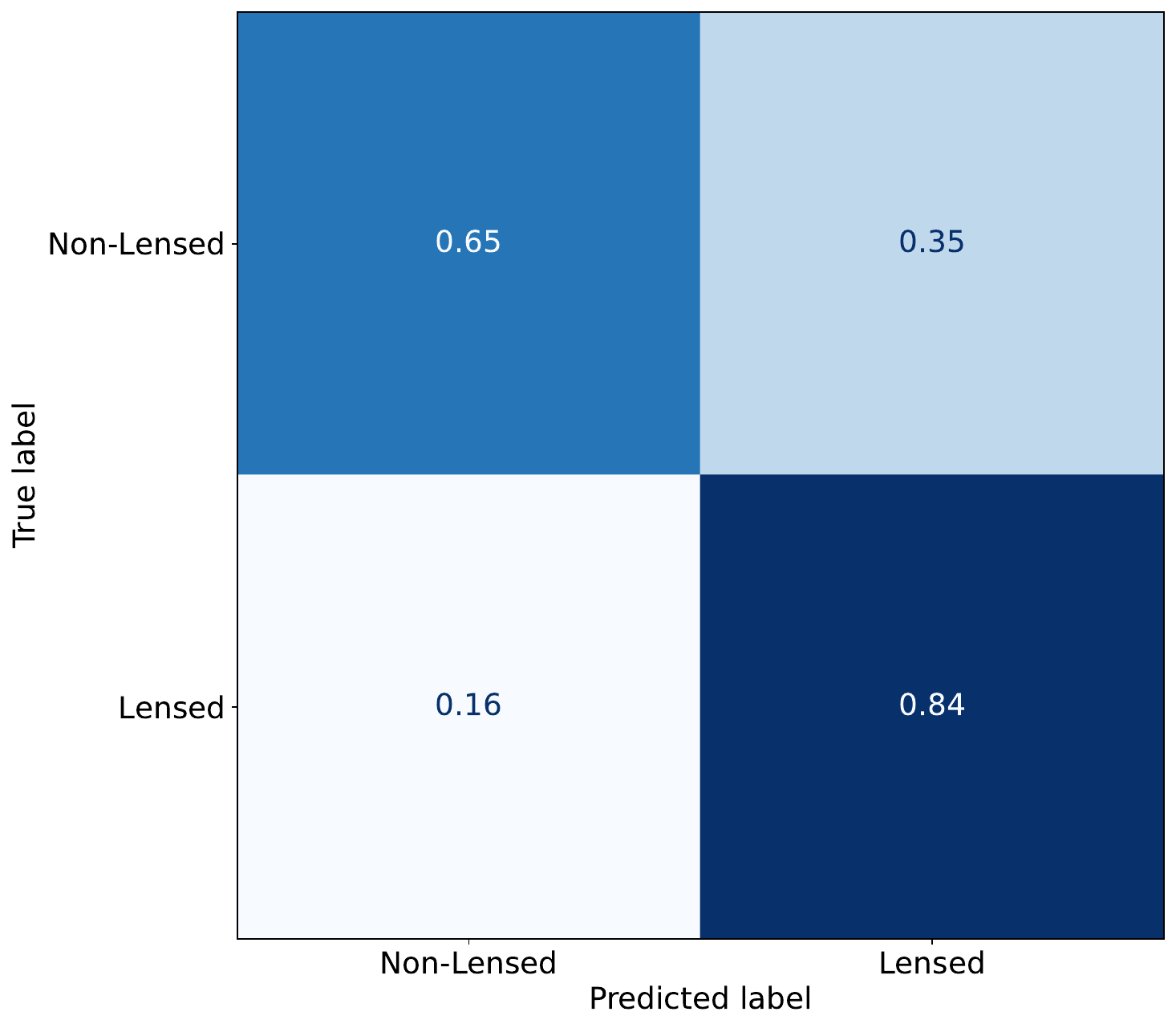}
        \caption{Logistic Regression}
        \label{fig:cm_logreg}
    \end{subfigure}
    \hfill
    \begin{subfigure}{0.32\textwidth}
                \includegraphics[width=\linewidth]{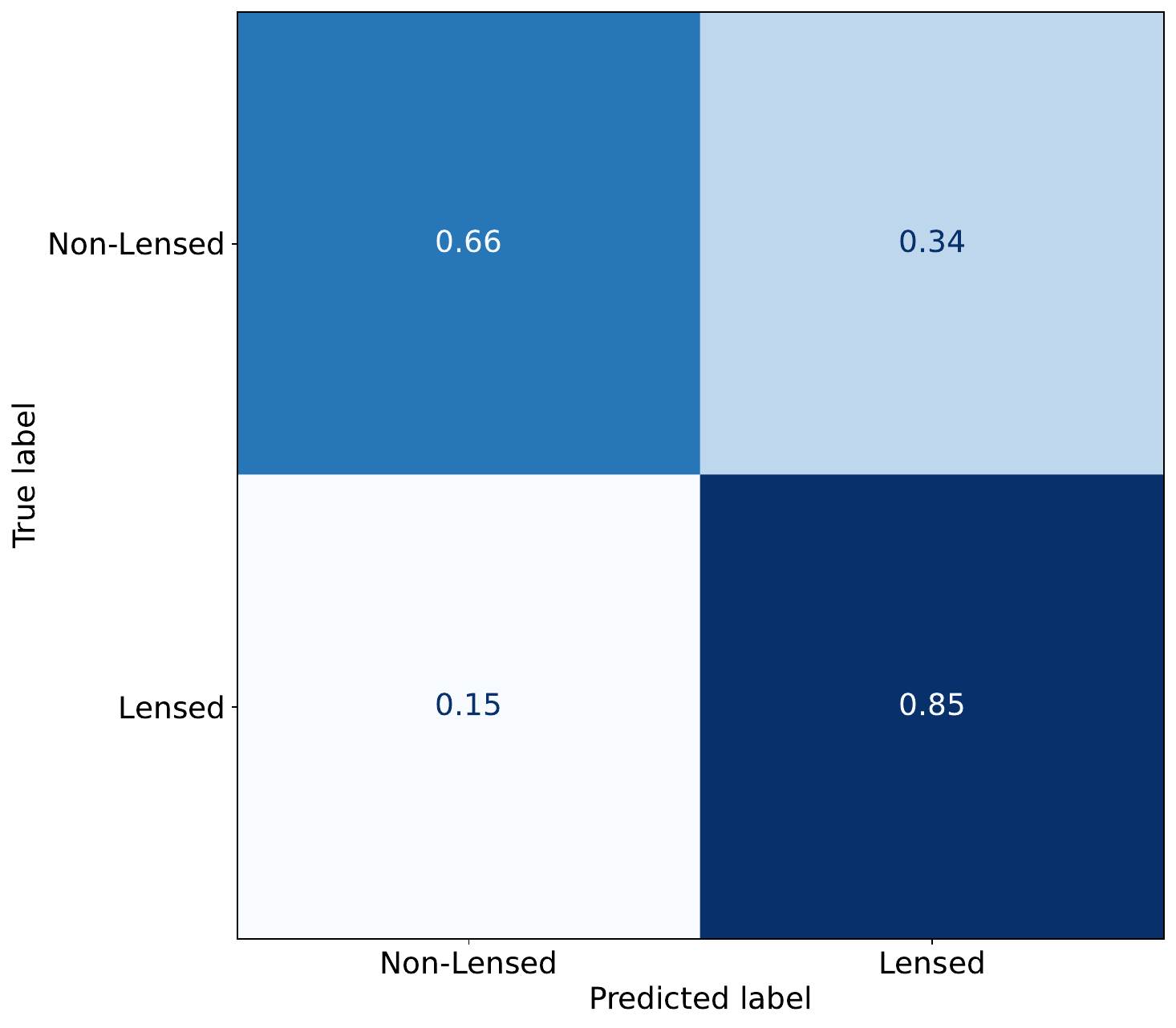}
        \caption{AdaBoost Classifier}
        \label{fig:cm_ada}
    \end{subfigure}

    \caption{Normalized confusion matrices for six machine learning models used to classify lensed and non-lensed GRBs.}
    \label{fig:conf_matrices}
\end{figure*}

To distinguish between micro-lensed and Non-Lensed gamma-ray bursts (GRBs), we employed multiple machine learning classifiers. The models were trained using a dataset of extracted features from simulated GRB light curves. The training process involved fitting each classifier to the training set $(X_{\text{train}}, y_{\text{train}})$, and the performance was evaluated on a separate test set $(X_{\text{test}}, y_{\text{test}})$.
The classifiers implemented include: Random Forest, Classifier (RF), Support Vector Machine (SVM), k-Nearest Neighbors (kNN), Bagging  Classifier, Logistic Regression, and AdaBoost Classifier (AdaBoost). Each classifier's performance was assessed using a confusion matrix and classification report. Figure~\ref{fig:conf_matrices} presents the normalized confusion matrices for all six models, providing insights into their classification effectiveness.

Furthermore, Table~\ref{tab:performance} summarizes the classification performance metrics, including precision, recall, and F1-score for each model.

\begin{figure}[htbp]
  
\centering    \includegraphics[width=0.5\textwidth]{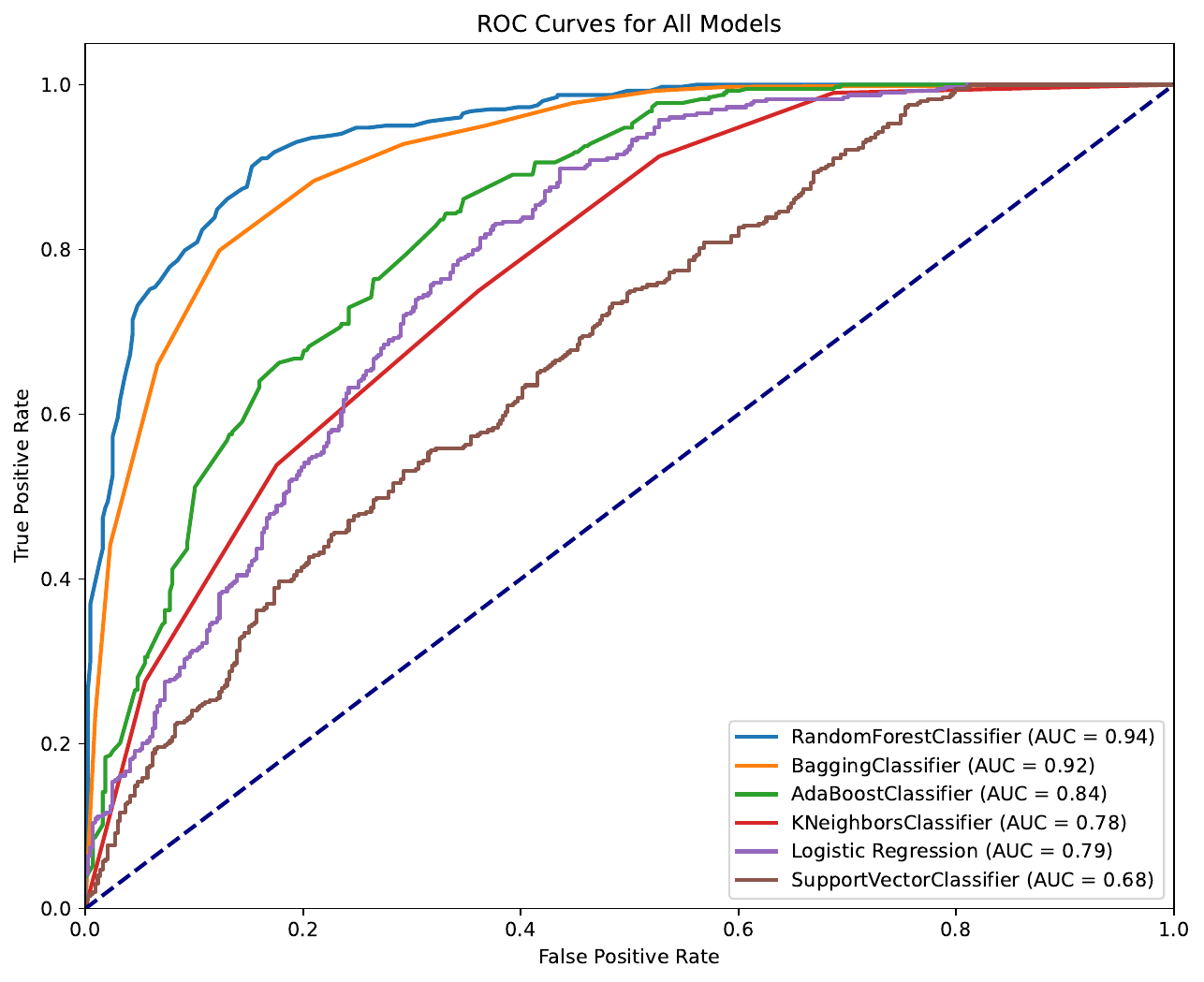} 
    \caption{ROC curves for all models. The area under the curve (AUC) for each model is indicated in the legend. The dashed blue line indicates a model with an AUC of 0.5, equivalent to random guessing with no discriminatory ability.}
    \label{fig:roc_curves}
\end{figure}

\begin{figure*}[ht!]
    \includegraphics[width=\textwidth]{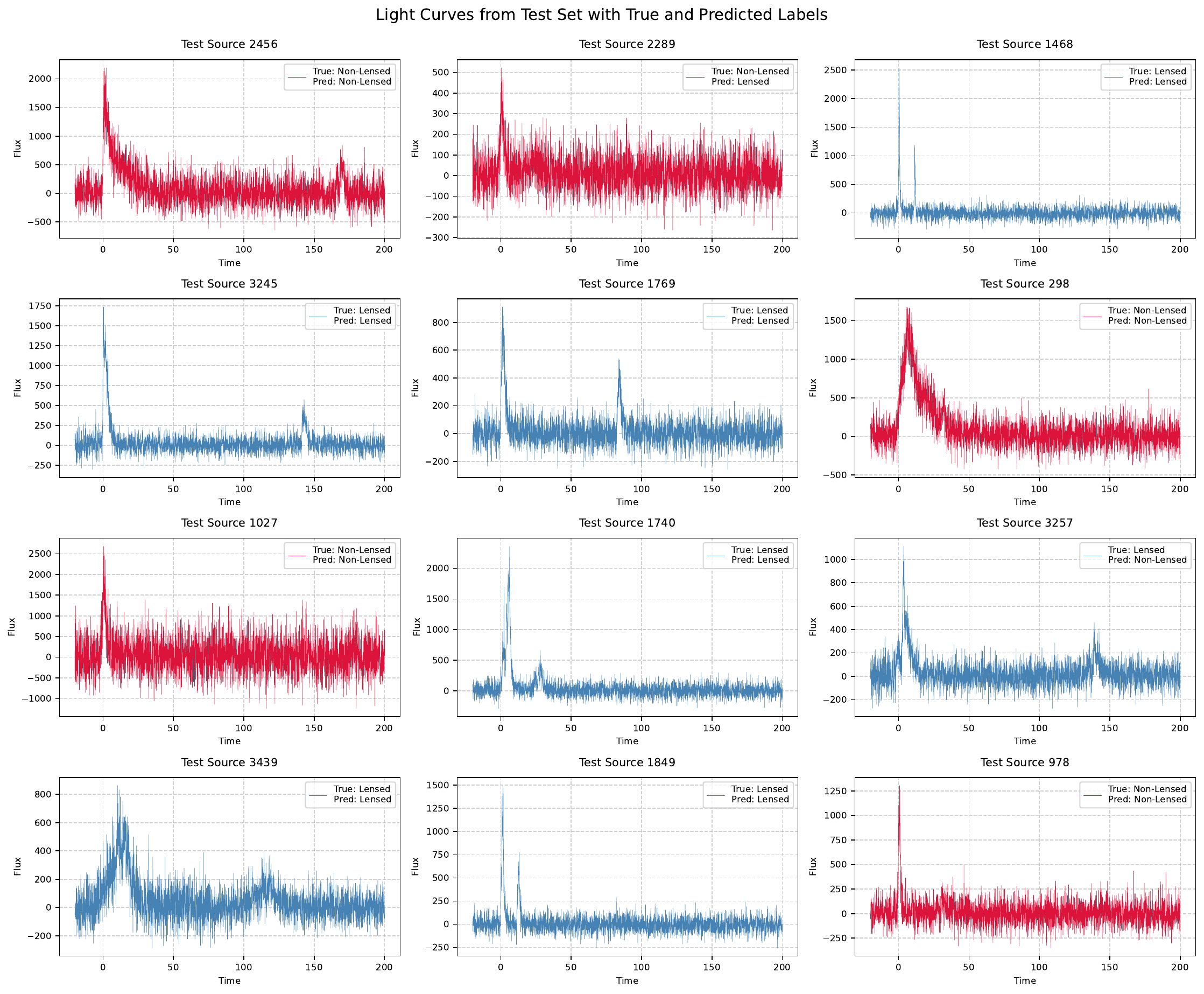}
    \caption{Light curves of selected sources from the test set. Each subplot shows the time series data for a specific source, with the true label (Lensed or Non-Lensed) and the predicted label from our machine learning model. The color coding distinguishes between Non-Lensed (red) and Lensed (blue) sources.}
    \label{fig:light_curves}
\end{figure*}

\begin{figure*}[ht!]
        \begin{subfigure}{0.32\textwidth}
                \includegraphics[width=\linewidth]{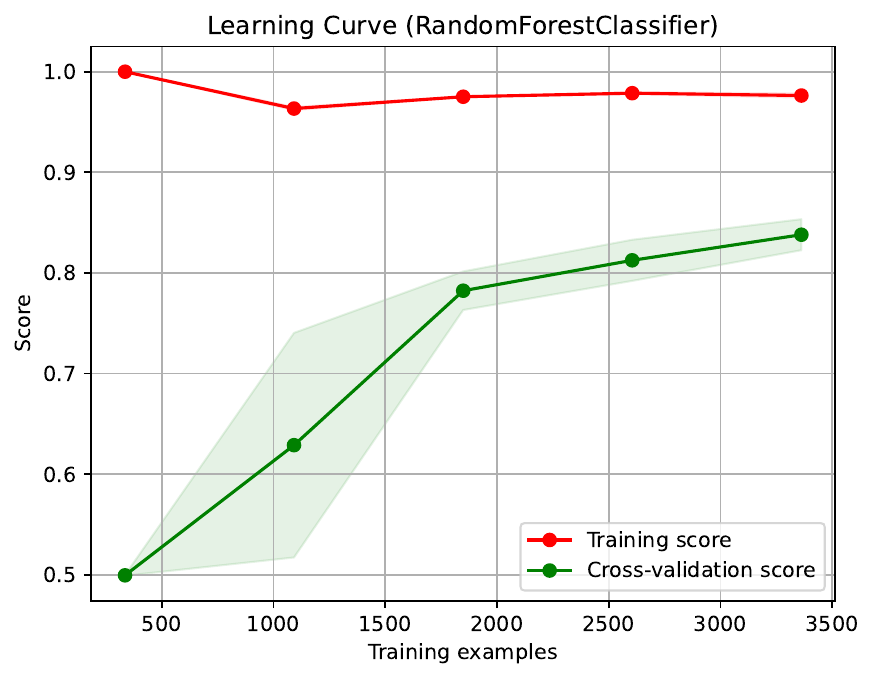}
        \label{fig:cm_rf}
    \end{subfigure}
    \hfill
   \begin{subfigure}{0.32\textwidth}
                \includegraphics[width=\linewidth]{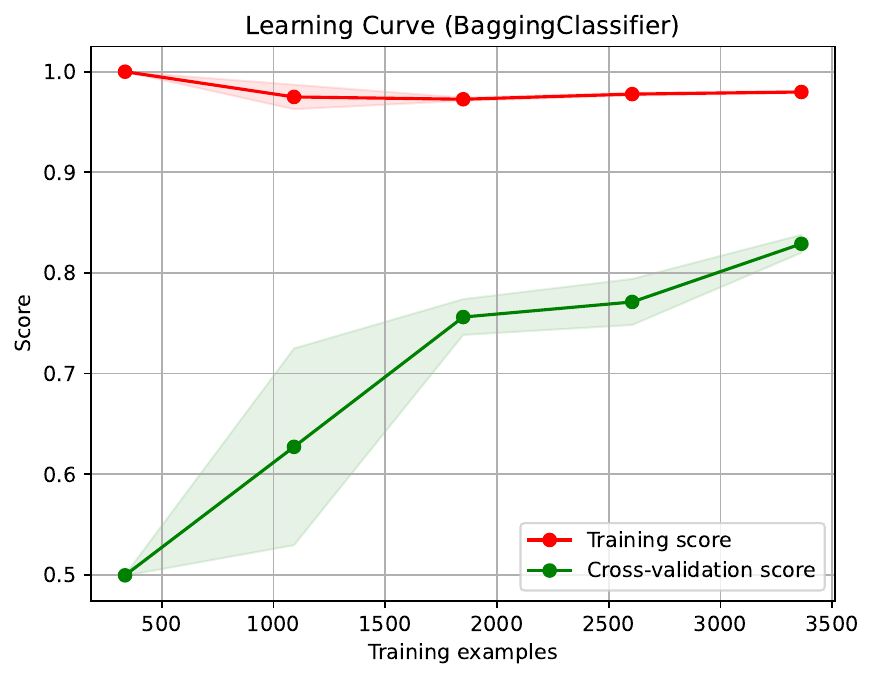}
        \label{fig:cm_bagging}
    \end{subfigure}
    \hfill
    \begin{subfigure}{0.32\textwidth}
                \includegraphics[width=\linewidth]{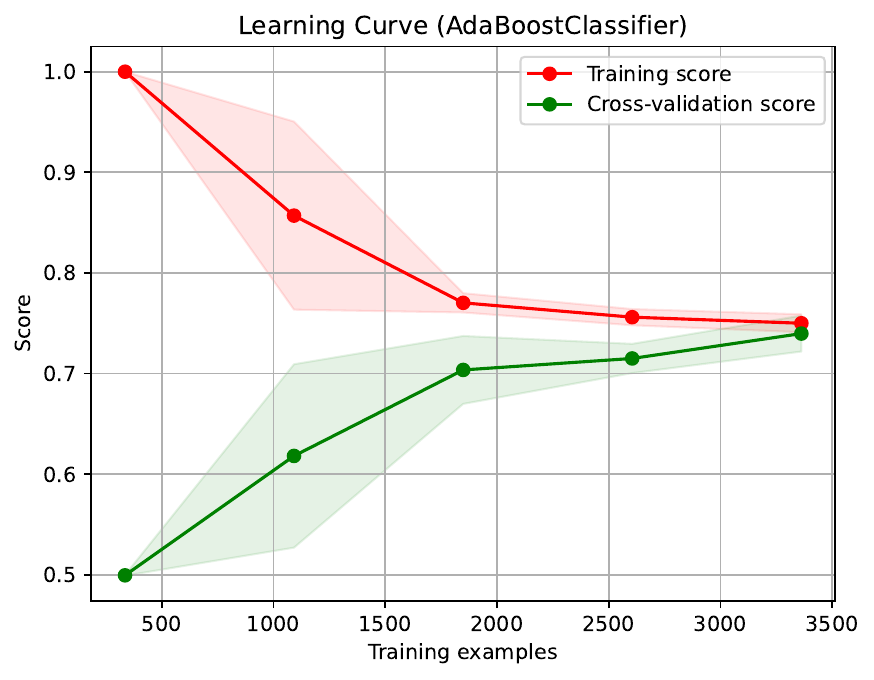}
        \label{fig:adaboost_curve}
    \end{subfigure}
    \vspace{0.5cm}
    \begin{subfigure}{0.32\textwidth}
                \includegraphics[width=\linewidth]{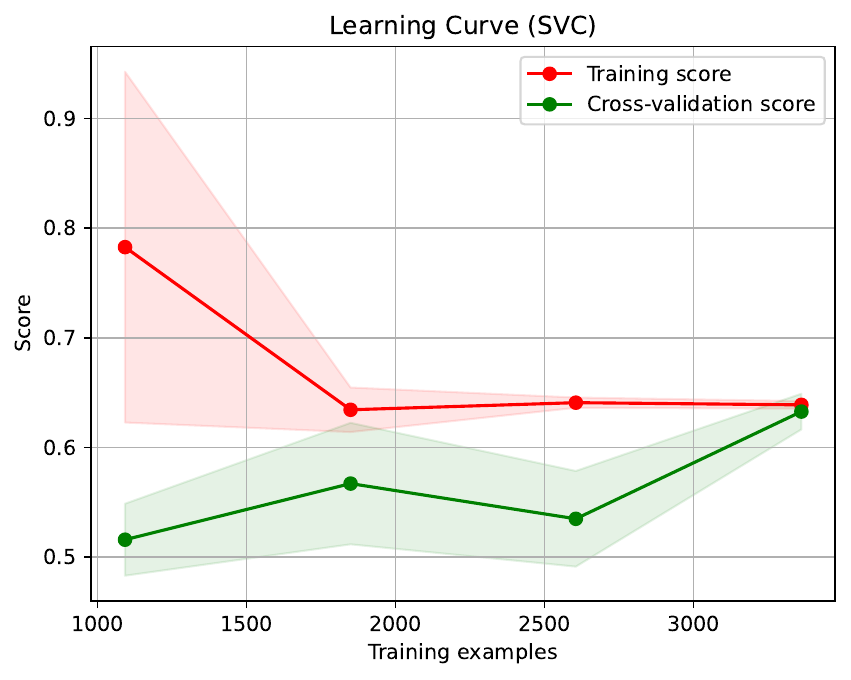}
        \label{fig:cm_svm}
    \end{subfigure}
    \hfill
    \begin{subfigure}{0.32\textwidth}
                \includegraphics[width=\linewidth]{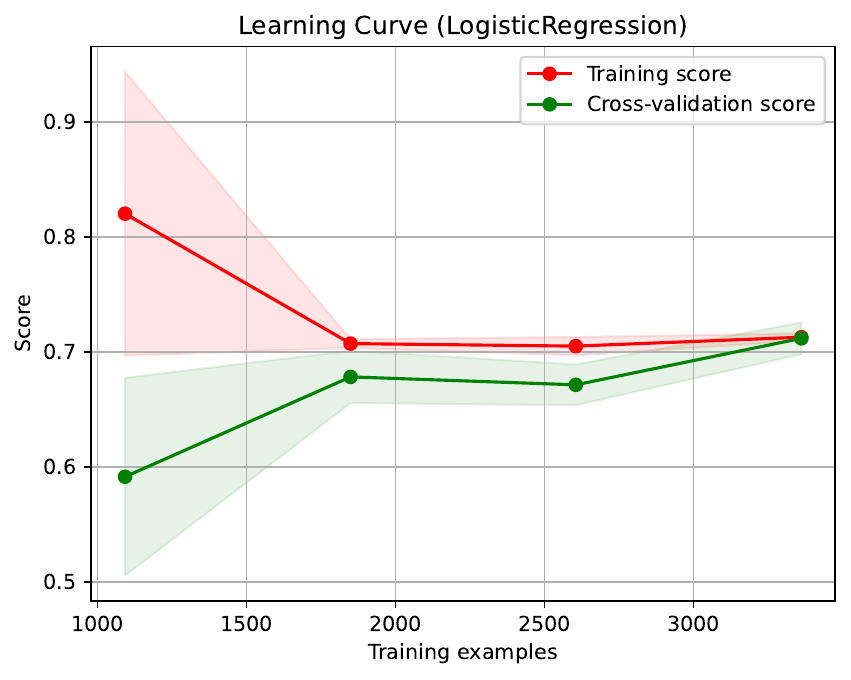}
        \label{fig:cm_mlp}
    \end{subfigure}
    \hfill
    \begin{subfigure}{0.32\textwidth}
                \includegraphics[width=\linewidth]{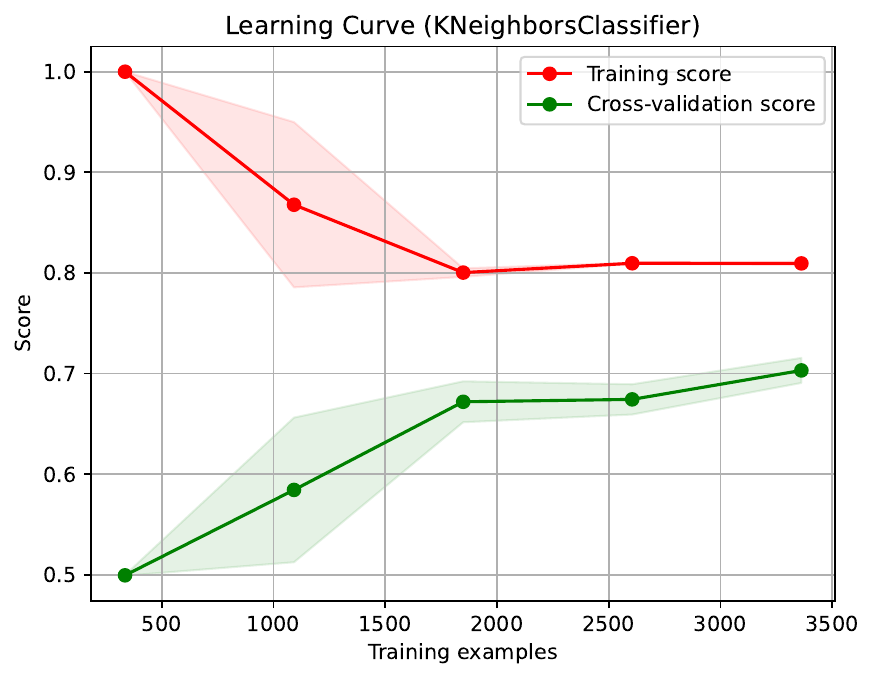}
        \label{fig:cm_knn}        

    \end{subfigure}

    \caption{Learning curve for our six models. The plot shows the training and cross-validation scores as the number of training examples increases. The shaded regions around the lines represent the variance of the training and cross-validation scores across different cross-validation folds.}
    \label{fig:learning_curves}
\end{figure*}

\section{Results}
\label{sec:results}

\subsection{Model Performance}
In this section, we present the performance of various machine learning models trained on the dataset for classifying Lensed and Non-Lensed objects. The evaluation metrics include precision, recall, F1-score, accuracy, and the Area Under the Curve (AUC) plotted in Figure \ref{fig:roc_curves}. The results are summarized in Table~\ref{tab:performance} and discussed in detail below.

The results indicate that the \textit{Random Forest (RF)} model outperforms all other models, achieving the highest accuracy of 86\% and an AUC of 0.94. It also demonstrates strong performance across both classes, with an F1-score of 0.86 for Non-Lensed objects and 0.87 for Lensed objects. This suggests that RF is robust and well-suited for this classification task.

The \textit{Bagging Classifier} also performs well, achieving an accuracy of 83\% and an AUC of 0.92 with an F1-score of 0.83 for Non-Lensed objects and 0.84 for Lensed objects. While slightly lower than RF, its performance is still competitive, indicating that ensemble methods are highly effective for this problem.

The \textit{k-Nearest Neighbors (kNN)} model achieves an accuracy of 79\% and an AUC of 0.78, with F1-scores of 0.77 for Non-Lensed objects and 0.80 for Lensed objects. 

The \textit{AdaBoostClassifier} achieves an accuracy of 75\% and an AUC of 0.84, with F1-scores of 0.73 for Non-Lensed objects and 0.76 for Lensed objects. Although its performance is lower than RF and Bagging, it still demonstrates reasonable capability in handling the classification task, particularly in terms of class separation.

The \textit{Logistic Regression} model achieves an accuracy of 74\% and an AUC of 0.79, with F1-scores of 0.72 for Non-Lensed objects and 0.76 for Lensed objects. While Logistic Regression performs better than SVM, its lower accuracy and F1-scores indicate that it may not be as effective as ensemble methods for this task. However, its AUC suggests that it has a reasonable ability to distinguish between the two classes.

Finally, the \textit{Support Vector Machine (SVM)} model achieves the lowest accuracy among the evaluated models, with an accuracy of 73\% and an AUC of 0.68. Its F1-scores of 0.69 for Non-Lensed objects and 0.76 for Lensed objects indicate that SVM may not be well-suited for this task, possibly due to the dataset's characteristics or the model's sensitivity to parameter tuning.

In Figure \ref{fig:light_curves}, we present the results of our analysis for Random Forest (the best model found in this study), focusing on the light curves of selected sources from the test set. These light curves are categorized based on their true labels (Lensed or Non-Lensed) and compared with the predicted labels from our machine learning model. The results demonstrate the effectiveness of our machine learning model in classifying light curves as Lensed or Non-Lensed. As shown in Figure~\ref{fig:light_curves}, the model correctly predicts the labels for most sources, with only a few misclassifications. For example, Source~2456 and Source~1468 are correctly identified as Non-Lensed and Lensed, respectively, while Source~2289 shows a slight discrepancy between the true and predicted labels.

\subsection{Learning Curve Analysis}
To evaluate the performance and generalization capability of different classifiers, we analyzed their learning curves. Figure \ref{fig:learning_curves} illustrates the learning curves of AdaBoost, Logistic Regression, Support Vector Classifier (SVC), k-Nearest Neighbors (kNN), Random Forest, and Bagging classifiers. Each curve shows the training and cross-validation scores as a function of the number of training examples.
These learning curves provide insights into how well each model learns from the data and generalizes to unseen examples.

\subsubsection{Random Forest and Bagging Classifiers}
Both the Random Forest and Bagging classifiers exhibit highly stable learning curves. The training score starts at 1.0, decreases slightly, and stabilises around 0.98. The validation score, initially at 0.5, steadily increases to 0.84, demonstrating generalisation performance. The small final gap of approximately 0.14 between the training and validation scores indicates that adding more data will lead to improved results. There is a slight indication of overfitting, which can be resolved with additional data. These models are the best-performing classifiers in this comparison.

\subsubsection{AdaBoost Classifier}
For the learning curve of the AdaBoost classifier, the training score starts at 1.0 and gradually decreases to approximately 0.75 as the number of training samples increases. Simultaneously, the cross-validation score improves from 0.5 to around 0.75, suggesting that increasing the training data will not enhance model generalization. Although the absence of a noticeable gap between training and validation scores suggests that the model is well-balanced and not overfitting, the overall performance remains relatively modest.

\subsubsection{For k-Nearest Neighbours (kNN)}
For kNN, the training score starts at 1.0, then decreases to about 0.80. Meanwhile, the validation score steadily rises from 0.5 to approximately 0.70. The final gap of about 0.1 between training and validation scores suggests that kNN slightly overfits the training data, but the improvement in validation performance indicates that it benefits from additional training examples.

\subsubsection{Logistic Regression}
 When it comes to Logistic Regression, the training score starts at about 0.82 and decreases with more training data, reaching 0.72 at the end. The validation score, however, increases from 0.60 to approximately 0.72.  The lack of a final gap between training and validation scores indicates reduced overfitting. The small final training and cross-validation scores indicate that the model is suffering from high bias. 

\subsubsection{Support Vector Classifier (SVC)}
The SVC learning curve follows a similar trend to Logistic Regression. The training score begins at approximately 0.78 and decreases as more data is introduced, to around 0.64. The validation score, initially at 0.52, gradually improves to 0.64. Large error bars at the beginning confirm the instability of the model with small datasets. While SVC improves with more data, its performance remains lower than ensemble methods, and the model suffers from high bias.

\subsection{Comparison of Models}
Based on the learning curves, the models can be compared as follows:

\begin{itemize}
    \item \textit{Best Generalization}: Random Forest and Bagging Classifier exhibit the best generalization, with small gaps between training and cross-validation scores. They are robust and perform well, showing no significant signs of underfitting. However, the small gap is a sign of slight overfitting, which can be resolved by additional training data.
    
    \item \textit{Moderate Generalization}:  AdaBoost classifier shows good generalization, with cross-validation scores close to training scores and minimal overfitting. However, its lower training and validation scores make it less attractive than Random Forest and Bagging Classifier for this task.
    
    \item \textit{Suboptimal Generalization}: The k-Nearest Neighbors (kNN) model exhibits slight overfitting, as indicated by the gap between training and cross-validation scores. In contrast, Logistic Regression and the Support Vector Classifier (SVC) show no evidence of overfitting; however, their uniformly low training and validation scores suggest underfitting or high bias.

\end{itemize}

\begin{table*}[t]
\begin{adjustbox}{max width=\textwidth}
\begin{tabular}{|l|c|c|c|c|c|c|c|c|}
\hline
\textbf{Model} & \multicolumn{3}{c|}{\textbf{Non-Lensed}} & \multicolumn{3}{c|}{\textbf{Lensed}} & \textbf{Accuracy} & \textbf{AUC} \\
\cline{2-7}
 & \textbf{Precision} & \textbf{Recall} & \textbf{F1-score} & \textbf{Precision} & \textbf{Recall} & \textbf{F1-score} & & \\
\hline
Random Forest (RF) & 0.93 & 0.79 & 0.86 & 0.81 & 0.94 & 0.87 & 86\% & 0.94 \\
Bagging Classifier & 0.88 & 0.79 & 0.83 & 0.79 & 0.88 & 0.84 & 83\% & 0.92 \\
k-Nearest Neighbors (kNN) & 0.89 & 0.67 & 0.77 & 0.72 & 0.91 & 0.80 & 79\% & 0.78 \\
AdaBoostClassifier & 0.82 & 0.66 & 0.73 & 0.69 & 0.85 & 0.76 & 75\% & 0.84 \\
Logistic Regression & 0.81 & 0.65 & 0.72 & 0.69 & 0.84 & 0.76 & 74\% & 0.79 \\
Support Vector Machine (SVM) & 0.86 & 0.58 & 0.69 & 0.66 & 0.90 & 0.76 & 73\% & 0.68 \\
\hline
\end{tabular}
\end{adjustbox}
\caption{Performance metrics for different machine learning models on the classification task. Precision, recall, and F1-score are reported for both Non-Lensed and Lensed classes, along with overall accuracy and AUC.}
\label{tab:performance}
\end{table*}

\subsection{Feature Importance Analysis}

To identify the most influential features contributing to the classification of gravitationally Lensed and Non-Lensed GRBs, we employed a Random Forest classifier with 100 trees, a maximum depth of 10, and specific constraints on node splitting to enhance generalization. After training the model on the extracted features, we computed feature importances using the Gini impurity-based ranking provided by the trained classifier.

\begin{figure}
\centering        \includegraphics[width=0.5\textwidth]{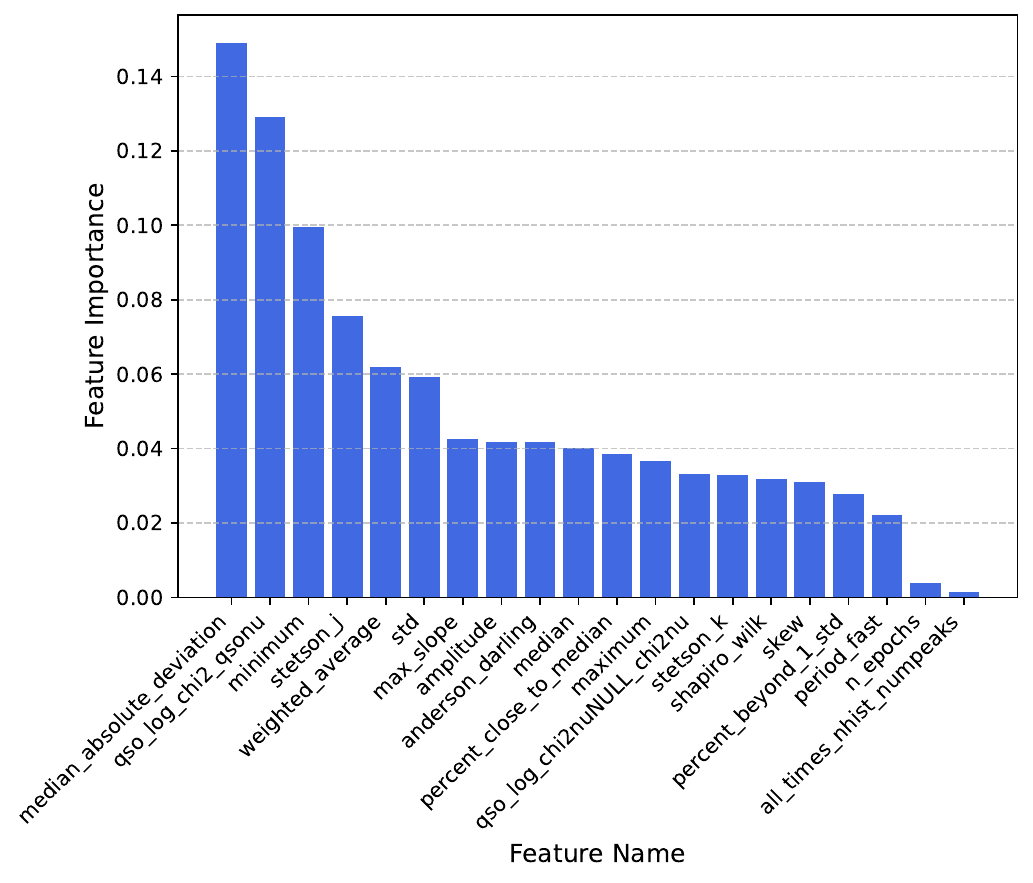}
    \caption{Top 20 most important features based on the trained Random Forest classifier. The x-axis represents the feature names, and the y-axis represents their importance scores. See section \ref{Feature_Selection}}
    \label{fig:feature_importance}
\end{figure}

Figure~\ref{fig:feature_importance} presents the most significant features based on their importance scores. The most influential features include  \texttt{median\_absolute\_deviation}, \texttt{qso\_log\_chi2\_qsonu}, \texttt{stetson\_j}, and \texttt{weighted\_average}, among others. These features capture key statistical and variability properties of light curves, which play a crucial role in distinguishing between Lensed and Non-Lensed GRBs.

The feature selection process revealed that time-series variability metrics such as \texttt{stetson\_j}, \texttt{stetson\_k}, and \texttt{shapiro\_wilk} contribute to classification performance, indicating that variability trends are pivotal in recognizing gravitational lensing signatures. Additionally, statistical descriptors like \texttt{median\_absolute\_deviation} and \texttt{skew} provide insights into the distributional properties of the light curves, further reinforcing their relevance to the classification task.

\subsection{Misclassified Cases}

\begin{figure}
        \includegraphics[width=1\textwidth]{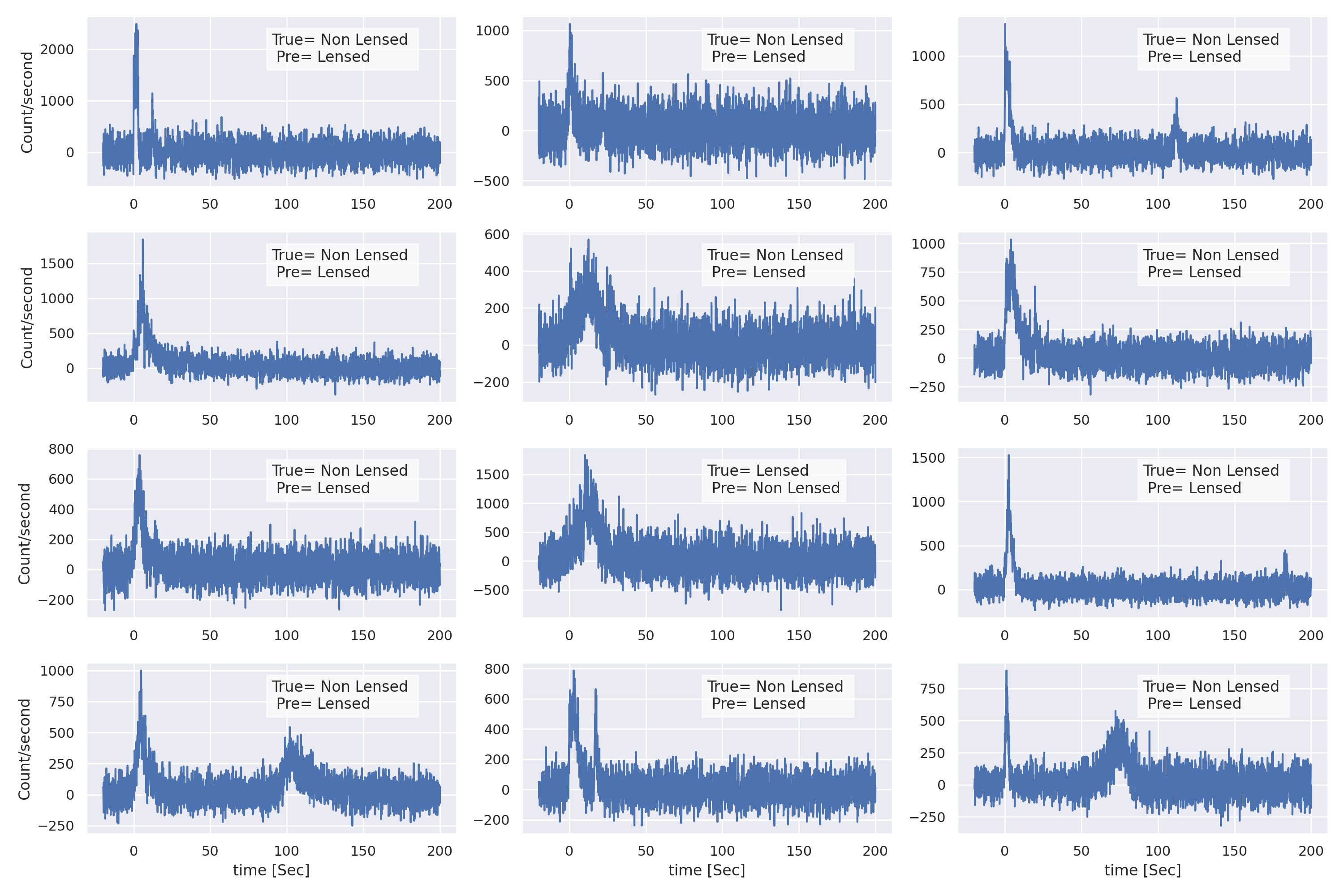}
    \caption{Misclassified cases into four failure modes:
First row: Similar Peaks from Separate Events; Morphological mimicry fools statistical tests.
Second row: Small Time Delays; Blended peaks ($\Delta t  \leq \tau$, where $\tau$ is the width of the first peak) obscure lensing signatures.
Third row: Low Magnification Ratios; Faint secondaries drown in noise.
Fourth row: Algorithmic False Detections; Noise/intrinsic variability misidentified as lensing.}
    \label{fig:misclass}
\end{figure}

Here we systematically analyze and categorize the misclassified cases, illustrated in Figure \ref{fig:misclass}. These instances fall into four distinct groups based on their failure modes:
\begin{enumerate}
    \item 
Similar Peaks from Separate Events (First row):
These Non-Lensed cases exhibit striking morphological similarity between peaks originating from distinct physical processes. Notably, even rigorous statistical tests like the  $\chi^2$ analysis \citep{Kalantari2} can falsely classify them as Lensed events.
    \item 
Small Time Delay Challenges (Second row):
Cases where the sub-threshold time delay between peaks ($\Delta t  \leq \tau$, where $\tau$ is the width of the first peak) causes peak blending. This temporal overlap creates degenerate light curve morphologies that obscure the lensing signature \citep{Kalantari}.
   \item 
Low Magnification Ratio Limitations (Third row):
Instances where the secondary peak's flux ratio renders it indistinguishable from background variations. The faint secondary emission becomes statistically undetectable given the instrumental signal-to-noise ratio \citep{Kalantari}.
    \item 
Algorithmic False Detections (Fourth row): The last row shows cases of misclassify due to false detection of our method. Boundary cases where our method incorrectly identifies variability patterns as Lensed, was expected as non of the trained models has 100\% accuracy and perfect $F_1$ score.
\end{enumerate}

\section{Application on observed GRBs}
\label{sec:application}

Up to this point, we have relied on simulated data to train the machine learning models, as the number of confirmed lensed GRBs remains extremely limited. Nevertheless, to assess the real-world applicability of our approach, we extracted features from six candidate lensed GRBs observed by Fermi, namely GRB081122520, GRB081126899, GRB090717034, GRB110517573, GRB200716957, and GRB210812699. The goal is to examine whether the trained models, which performed well on simulated data, could correctly identify these real candidates as lensed GRBs.
The results, summarized in Table \ref{tab:model_predictions}, show that different classifiers yield varied levels of success. For example, the AdaBoostClassifier correctly identified five out of six candidates as lensed, while the RandomForestClassifier predicted none as lensed. This inconsistency reflects the challenges posed by the simulation-to-reality gap, as the models were trained solely on simulated data, which may not fully capture the complexity of real GRB light curves.

\begin{table}[ht]
\small  
\setlength{\tabcolsep}{4pt}  
\caption{Predictions of different machine learning models for six observed candidate lensed GRBs. A value of 1 indicates a prediction of Lensed GRB, while 0 indicates Non-Lensed.}
\label{tab:model_predictions}
\begin{tabular}{lcccccc}
\hline
\textbf{Model} & \textbf{081122520} & \textbf{081126899} & \textbf{090717034} & \textbf{110517573} & \textbf{200716957} & \textbf{210812699} \\
\hline
BaggingClassifier       & 0 & 0 & 0 & 1 & 1 & 0 \\
AdaBoostClassifier      & 1 & 0 & 1 & 1 & 1 & 1 \\
KNeighborsClassifier    & 0 & 0 & 0 & 1 & 1 & 1 \\
SupportVectorClassifier & 1 & 0 & 0 & 0 & 1 & 1 \\
RandomForestClassifier  & 0 & 0 & 0 & 0 & 0 & 0 \\
Logistic Regression     & 0 & 0 & 1 & 0 & 1 & 0 \\
\hline
\end{tabular}
\end{table}

\begin{figure}
        \includegraphics[width=1\textwidth]{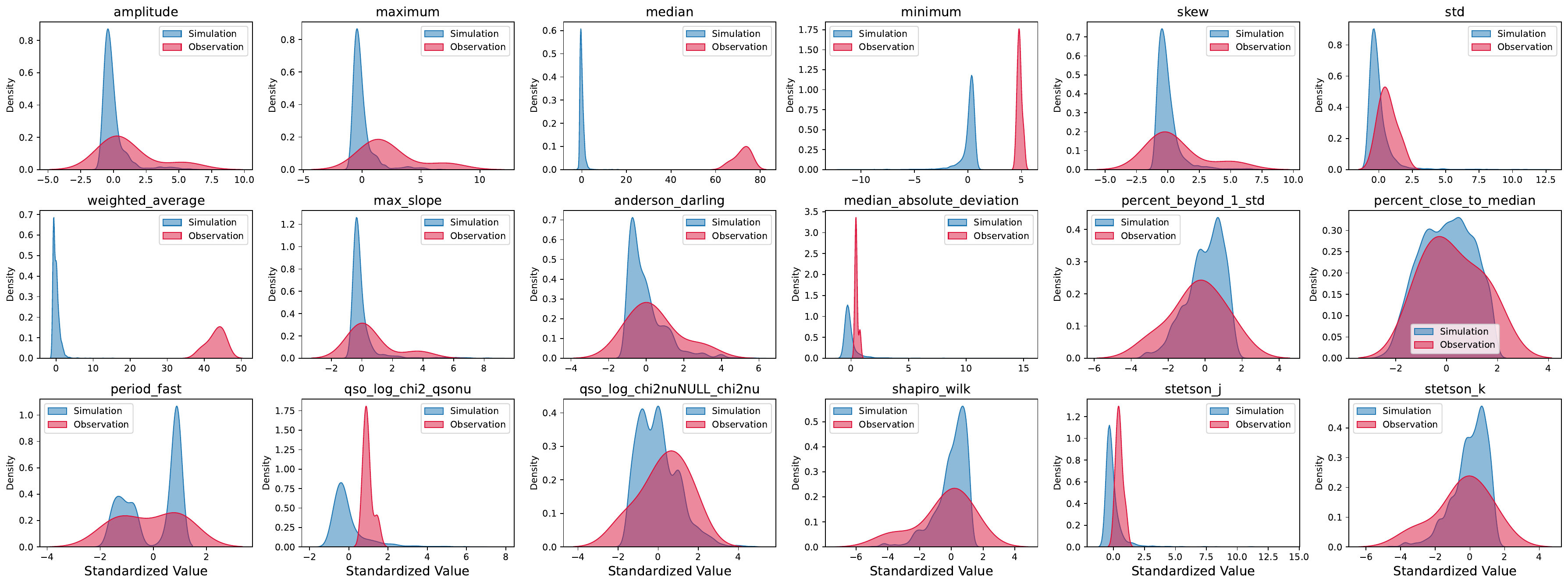}
    \caption{Kernel Density Estimates (KDEs) comparing the distributions of simulated (blue) and observed (orange) features after standardisation (mean = 0, standard deviation = 1). Each subplot represents a distinct feature, with the shaded area indicating the probability density.}
    \label{fig:feature_dist}
\end{figure}

To further understand this behavior, we conducted a comparative analysis of the feature distributions between real GRB data and our simulations. As can be seen from Figure \ref{fig:feature_dist}, we found noticeable separation between the feature distribution of the two domains, especially for the most important features like \texttt{median\_absolute\_deviation}, \texttt{stetson\_j}, \texttt{qso\_log\_chi2\_qsonu} , and \texttt{weighted\_average}.

To evaluate the similarity between simulated and real GRB light curve feature distributions, we performed the Kolmogorov–Smirnov (K–S) test \citep{Kolmogorov1933, Smirnov1948} on each extracted feature. This statistical test assesses whether two samples are drawn from the same distribution. In practice, a large K–S statistic with a low p-value suggests strong differences between simulated and real GRB feature distributions, while a small statistic with a high p-value suggests close agreement.
The results (Table \ref{tab:ks_test}) indicate that several features, such as \texttt{maximum, median, minimum, weighted\_average, median\_absolute\_deviation, qso\_log\_chi2\_qsonu}, and \texttt{stetson\_j} show statistically significant differences between the two domains, with p-values less than 0.05. This divergence highlights the presence of a simulation-to-reality gap, explaining the limited performance of some models when applied to real observed GRBs. On the other hand, some features, such as \texttt{amplitude, skew, n\_epochs}, and \texttt{all\_times\_nhist\_numpeaks}, showed no significant distributional differences, suggesting that they are more robust for real-world generalization. These findings suggest that the simulated light curves, while effective for training, do not fully capture the complexity, noise characteristics, and variability present in real observational data. As a result, the model’s reduced performance on real candidate GRBs is an expected consequence of domain shift.

\begin{table}[ht]
\caption{Results of the Kolmogorov-Smirnov (K-S) test comparing feature distributions between simulated and real GRBs. K–S Statistic is the maximum distance between the empirical cumulative distribution functions (ECDFs) of the two samples, which take values ranging from 0 (perfect match) to 1 (complete mismatch).}
\label{tab:ks_test}
\begin{tabular}{lcc}
\hline
\textbf{Feature} & \textbf{K-S Statistic} & \textbf{p-value} \\
\hline
amplitude & 0.477 & 0.0891 \\
percent\_beyond\_1\_std & 0.349 & 0.3708 \\
maximum & {0.872} & {0.0000} \\
median & {1.000} & {0.0000} \\
minimum & {1.000} & {0.0000} \\
skew & 0.222 & 0.8719 \\
std & {0.610} & {0.0114} \\
weighted\_average & {1.000} & {0.0000} \\
max\_slope & {0.650} & {0.0053} \\
anderson\_darling & 0.353 & 0.3576 \\
median\_absolute\_deviation & {0.878} & {0.0000} \\
percent\_close\_to\_median & 0.239 & 0.8138 \\
period\_fast & 0.208 & 0.9133 \\
qso\_log\_chi2\_qsonu & {0.864} & {0.0000} \\
qso\_log\_chi2nuNULL\_chi2nu & 0.376 & 0.2875 \\
shapiro\_wilk & 0.232 & 0.8391 \\
stetson\_j & {0.790} & {0.0002} \\
stetson\_k & 0.257 & 0.7401 \\
all\_times\_nhist\_numpeaks & 0.021 & 1.0000 \\
n\_epochs & 0.054 & 1.0000 \\
\hline
\end{tabular}
\end{table}

\section{Summary and Conclusions}
We investigated the possibility of classifying micro-lensed and Non-Lensed GRBs through the application of some popular machine learning models. The method turned out to be successful and can be applicable for real-time observations.
Overall, the results demonstrate that ensemble methods such as Random Forest and Bagging Classifier are the most effective for this classification task, achieving high accuracy and balanced performance across both classes, and very good AUC values. In contrast, models like SVC, Logistic Regression, and kNN struggle to generalize well, highlighting the importance of model selection for this problem. The AUC metric further reinforces the superiority of ensemble methods, as they exhibit the highest ability to distinguish between Lensed and Non-Lensed objects.

Additionally, analysing the learning curve of the six models demonstrates that the Random Forest and Bagging classifiers demonstrate the best trade-off between training and validation performance, with minimal overfitting and the highest validation scores. AdaBoost also generalizes well but has a less satisfactory validation score. Logistic Regression and SVC show a steady improvement in validation performance, though their lower final scores suggest they may not be the best choices for this dataset. The kNN classifier, while improving with additional data, exhibits signs of slight overfitting and remains behind the ensemble methods in performance. 

Finally, testing on six candidate lensed GRBs observed by Fermi revealed that while some classifiers (e.g., AdaBoost) showed partial success, others were less effective, highlighting a clear simulation-to-reality gap. The Kolmogorov–Smirnov test confirmed significant distributional differences in several features, which explains the reduced performance. However, a subset of features (e.g., amplitude, skew, number of epochs) proved more robust, suggesting that future work should refine simulations and prioritize such stable features to enhance real-world generalization.

\label{conclusion}

\bibliographystyle{plainnat} 
\bibliography{references_arxive.bib}{}

\end{document}